\documentstyle{mn} 

\newif\ifAMStwofonts

\newcommand{\cooa}[3]{\makebox[3mm][r]{#1}\hspace{2mm}\makebox[3mm][r]{#2}
                      \hspace{2mm}\makebox[4.5mm][r]{#3}}

\newcommand{\cood}[3]{\,\makebox[3mm][r]{#1}\hspace{2mm}\makebox[3mm][r]{#2}
                      \hspace{1mm}\makebox[3mm][r]{#3}}

\newcommand{\mdot}[3]{\makebox[10mm][r]{#1}%
\makebox[10mm][l]{\,$^{{+}#2}_{{-}#3}$}}

\newcommand{\mdotvoid}[3]{\makebox[10mm][r]{#1}%
\makebox[10mm][l]{\,$^{   #2}_{   #3}$}}

\newcommand{\rcool}[3]{\makebox[6mm][r]{#1}%
\makebox[6mm][l]{\,$^{{+}#2}_{{-}#3}$}}

\newcommand{\ldotss}{\mbox{}\hspace{3mm}\ldots}

\newcommand{\reduceme}{\mbox{R\raisebox{-0.35ex}{E}D%
\hspace{-0.05em}\raisebox{0.85ex}{uc}\hspace{-0.90em}%
\raisebox{-.35ex}{{m}}\hspace{0.05em}E}}

\newcommand{\namelarge}[1]{\raisebox{0.ex}[1.5ex][2.5ex]{#1}}

\ifoldfss
  \ifCUPmtlplainloaded \else
    \NewTextAlphabet{textbfit} {cmbxti10} {}
    \NewTextAlphabet{textbfss} {cmssbx10} {}
    \NewMathAlphabet{mathbfit} {cmbxti10} {} 
    \NewMathAlphabet{mathbfss} {cmssbx10} {} 
  \fi
  \ifAMStwofonts
    \ifCUPmtlplainloaded \else
      \NewSymbolFont{upmath} {eurm10}
      \NewSymbolFont{AMSa} {msam10}
      \NewMathSymbol{\upi}     {0}{upmath}{19}
      \NewMathSymbol{\umu}     {0}{upmath}{16}
      \NewMathSymbol{\upartial}{0}{upmath}{40}
      \NewMathSymbol{\leqslant}{3}{AMSa}{36}
      \NewMathSymbol{\geqslant}{3}{AMSa}{3E}

      \let\leq=\leqslant 
       
    \fi
  \fi
\fi 

\ifnfssone
  \newmathalphabet{\mathit}
  \addtoversion{normal}{\mathit}{cmr}{m}{it}
  \addtoversion{bold}{\mathit}{cmr}{bx}{it}
  \newmathalphabet{\mathbfit} 
  \addtoversion{normal}{\mathbfit}{cmr}{bx}{it}
  \addtoversion{bold}{\mathbfit}{cmr}{bx}{it}
  \newmathalphabet{\mathbfss} 
  \addtoversion{normal}{\mathbfss}{cmss}{bx}{n}
  \addtoversion{bold}{\mathbfss}{cmss}{bx}{n}
  \ifAMStwofonts
    \ifCUPmtlplainloaded \else
      \UseAMStwoboldmath
      \makeatletter
      \new@mathgroup\upmath@group
      \define@mathgroup\mv@normal\upmath@group{eur}{m}{n}
      \define@mathgroup\mv@bold\upmath@group{eur}{b}{n}
      \edef\UPM{\hexnumber\upmath@group}
      \new@mathgroup\amsa@group
      \define@mathgroup\mv@normal\amsa@group{msa}{m}{n}
      \define@mathgroup\mv@bold\amsa@group{msa}{m}{n}
      \edef\AMSa{\hexnumber\amsa@group}
      \makeatother
      \mathchardef\upi="0\UPM19
      \mathchardef\umu="0\UPM16
      \mathchardef\upartial="0\UPM40
      \mathchardef\leqslant="3\AMSa36
      \mathchardef\geqslant="3\AMSa3E

      \let\leq=\leqslant 

    \fi
  \fi
\fi 

\ifnfsstwo
  \DeclareMathAlphabet{\mathbfit}{OT1}{cmr}{bx}{it}
  \SetMathAlphabet\mathbfit{bold}{OT1}{cmr}{bx}{it}
  \DeclareMathAlphabet{\mathbfss}{OT1}{cmss}{bx}{n}
  \SetMathAlphabet\mathbfss{bold}{OT1}{cmss}{bx}{n}
  \ifAMStwofonts
    \ifCUPmtlplainloaded \else
      \DeclareSymbolFont{UPM}{U}{eur}{m}{n}
      \SetSymbolFont{UPM}{bold}{U}{eur}{b}{n}
      \DeclareSymbolFont{AMSa}{U}{msa}{m}{n}
      \DeclareMathSymbol{\upi}{0}{UPM}{"19}
      \DeclareMathSymbol{\umu}{0}{UPM}{"16}
      \DeclareMathSymbol{\upartial}{0}{UPM}{"40}
      \DeclareMathSymbol{\leqslant}{3}{AMSa}{"36}
      \DeclareMathSymbol{\geqslant}{3}{AMSa}{"3E}

      \let\leq=\leqslant 

    \fi
  \fi
\fi 

\ifCUPmtlplainloaded \else
  \ifAMStwofonts \else 
    \def\upi{\pi}
    \def\umu{\mu}
    \def\upartial{\partial}
  \fi
\fi

\title[Spectral gradients in central cluster galaxies]
{Spectral gradients in central cluster galaxies: \\
further evidence of star formation in cooling flows}
\author[N.~Cardiel, J.~Gorgas and A.~Arag\'{o}n-Salamanca]
       {N.~Cardiel,$^1$ J.~Gorgas,$^1$ and A.~Arag\'{o}n-Salamanca$^2$\\
        $^1$Departamento de Astrof\'{\i}sica, Facultad de Ciencias
F\'{\i}sicas, Universidad Complutense de Madrid, 28040 Madrid, Spain\\
        $^2$Institute of Astronomy, Madingley Road, Cambridge, CB3 0HA}
\date{Accepted 1988 December 15.
      Received 1988 December 14;
      in original form 1988 October 11}

\pagerange{\pageref{firstpage}--\pageref{lastpage}}
\pubyear{1997}

\begin{document}

\maketitle

\label{firstpage}

\begin{abstract}
We have obtained radial gradients in the spectral features
\mbox{$\lambda$4000-\AA} break (D$_{4000}$) and Mg$_2$ for a sample of
11 central cluster galaxies (CCGs): 8 in clusters with cooling flows
and 3 in clusters without. After careful removal of the emission lines
found within the D$_{4000}$ and the Mg$_2$ bandpasses for some
objects, the new data strongly confirm the correlations between
line-strength indices and the cooling flow phenomenon found in our
earlier study (Cardiel, Gorgas \& Arag\'{o}n-Salamanca 1995).  We find
that such correlations depend on the presence and characteristics of
emission lines in the inner regions of the CCGs. The nuclear indices
are correlated with the mass deposition rate ($\dot{M}$) {\it only\/}
when emission lines are found in the central regions of the galaxies.
The central D$_{4000}$ and Mg$_2$ indices in cooling flow galaxies {\it
without\/} emission lines are completely consistent with the indices
measured in CCGs in clusters without cooling flows.  CCGs in cooling
flow clusters exhibit a clear sequence in the \mbox{D$_{4000}$--Mg$_2$}
plane, with a neat segregation depending on emission-line types
(Heckman et al.  1989) and blue morphology (McNamara 1997). This
sequence can be modelled, using stellar population models with a normal
IMF, by a recent ($\sim 0.1$~Gyr old) burst of star formation, although
model uncertainties do not allow us to completely discard a continuous
star formation or a series of bursts over the last few Gyr.  In CCGs
with emission lines, the gradients in the spectral indices are flat or
positive inside the emission-line regions, suggesting the presence of
young stars.  Outside the emission-line regions, and in cooling flow
galaxies without emission lines, gradients are negative and consistent
with those measured in CCGs in clusters without cooling flows and giant
elliptical galaxies.  Index gradients measured exclusively in the
emission-line region correlate with $\dot{M}$. Using the same
population models we have estimated the radial profiles of the
mass transformed into new stars.  The derived profiles are remarkably
parallel to the expected radial behaviour of the mass deposition rate
derived from X-ray observations.  Moreover, a large fraction ({\it
probably most\/}) of the cooling flow gas accreted into the
emission-line region is converted into stars.  In the light of this new
data, we discuss the evolutionary sequence suggested by McNamara
(1997), in which radio triggered star formation bursts take place
several times during the lifetime of the cooling flow. We conclude that
this scenario is consistent with the available observations.
\end{abstract}

\section{Introduction}

The cooling time of the hot intra cluster medium is shorter than the
Hubble time in the central regions of most galaxy clusters (e.g. Arnaud
1988, Edge, Stewart \& Fabian 1992), which suggests that large amounts
of cool material should be settling at the cluster centre.  This
process, normally known as a {\it cooling flow\/}, may take place at
rates of \mbox{$\dot{M} \sim 100$~M$_{\odot}$~yr$^{-1}$} (Fabian 1994),
which implies that \mbox{$M_{\rm Total} \sim 10^{12}$~M$_{\odot}$} of
gas would cool and accrete onto the central cluster elliptical galaxies
over the lifetime of the clusters. Considering the idealised picture of
an homogeneous cooling flow, all the material will be deposited in the
centre of the cluster. However, X-ray data show that the gas is highly
inhomogeneous and, very likely, most of the gas must be cooling out at
large radii, whilst only a fraction reaches the centre of the flow.  In
fact, mass deposition profiles are reasonably fitted by $\dot{M}(<r)
\propto r$ (Thomas, Fabian \& Nulsen 1987), although some clusters do
not follow this trend beyond a certain radius $r \ga 60$~kpc (David,
Jones \& Forman 1994; Irwin \& Sarazin 1995).  For comprehensive
reviews on cooling flows see Fabian, Nulsen \& Canizares (1984, 1991),
Sarazin (1986) and Fabian (1994).

A question of open debate is the nature of the final repository of the
cool gas. The discovery of significant excess absorption in the soft X-ray
spectra of some galaxy clusters containing cooling flows (White et al.
1991; Johnstone et al. 1992; Miyaji et al. 1993; Allen et al. 1993,
1995; Fabian et al. 1994; Irwin \& Sarazin 1995; Allen \& Fabian 1997a,
1997b) requires \mbox{$\sim 10^{11}$--$10^{12}$~M$_{\odot}$} of
absorbing cold gas, probably in small, pressure-confined, magnetised
clouds (Daines, Fabian \& Thomas 1994). This value is comparable to
$M_{\rm Total}$, the mass deposition rate integrated over the age of
the clusters. Fabian, Johnstone \& Daines (1994) have postulated that
dust can form in the cold clouds embedded in cooling flows, which would
explain why the clouds are almost undetectable outside the X-ray
wavelength range. However, there is a great uncertainty concerning the
nature of the cold material producing this absorption (Sarazin 1997,
Laor 1997).  Voit \& Donahue (1995), from the limits imposed by 21~cm
and CO observations (McNamara, Bregman \& O'Connell 1990; Antonucci \&
Barvainis 1994; McNamara \& Jaffe 1994; O'Dea et al. 1994; O'Dea,
Gallimore \& Baum 1995), suggest that it is unlikely that significant
amounts of cold gas can remain undetectable.  On the other hand, very
recently Jaffe \& Bremer (1997), using \mbox{$K$-band} spectroscopy,
have found that the inner few kpc of central cluster galaxies (CCGs) in
cooling flows exhibit strong emission in the \mbox{H$_2$(1--0)S(1)
line}, which is not seen in a comparable sample of non cooling flow
galaxies.  These authors suggest that it is likely that a large mass of
additional molecular material covers the cooling flow as a whole,
producing the soft X-ray absorption.

Since it seems clear that at least a fraction of the cool gas must
reach the inner regions of the CCGs, it is reasonable to think that
part of this gas could accumulate as molecular clouds which,
presumably, would form stars.  Unusual diffuse blue light and colours
(Sarazin \& O'Connell 1983; Bertola et al. 1986; Johnstone, Fabian \&
Nulsen 1987 ---hereafter JFN87---; Romanishin 1987; Crawford et al.
1989; McNamara \& O'Connell 1989, 1992 ---hereafter MO89, MO92---,
1993; Thuan \& Puschell 1989; N{\o}rgaard-Nielsen, J{\o}rgensen \&
Hansen 1990; Allen et al.  1992; Crawford \& Fabian 1993; Crawford et
al.  1995; Allen 1995; McNamara 1995; Hansen, J{\o}rgensen \&
N{\o}rgaard-Nielsen 1995; Cardiel, Gorgas \& Arag\'{o}n-Salamanca 1995
---hereafter CGA95---; Melnik, Gopal-Krishna \& Terlevich 1997), strong
low-ionization optical line emission (Baum 1992, Donahue \& Voit 1997,
and references therein), and powerful radio sources (see Jaffe 1992,
Burns et al.  1997, for reviews) are usually associated with the inner
few kpc of central dominant galaxies in clusters with clear X-ray
signatures of cooling flows.  Although with a large scatter, the
strength of the optical anomalies are correlated with mass deposition
rates (JFN87; MO89; MO92; Heckman et al.  1989 ---hereafter HBvBM---;
CGA95; McNamara 1997). In most cases star formation has been invoked to
reproduce the excess blue light in the continuum, with typical derived
star formation rates ranging from a few to some tens of solar masses
per year. It is important to stress that these values normally only
account for \mbox{$\la 10$} per cent of the total X-ray derived mass
deposition rates.

McNamara \& O'Connell (1993) found that the blue colours observed in
A~1795 and A~2597 are located in patches that are coincident with the
radio lobes of the central galaxies, which suggests a connection between the
radio source and the star formation (or, at least, the source of the blue
light). Although power-law spectra could also be an alternative explanation
for the origin of the excess blue light (e.g. Crawford \& Fabian 1993), the
absence of polarization in \mbox{$U$-band} 
observations of the blue lobes in
A~1795 (McNamara et al. 1996b) makes this scenario unlikely. In addition,
observations of this object with the {\it Hubble Space Telescope\/} (McNamara
et al. 1996a; Pinkney et al. 1996), and the {\it Ultraviolet Imaging
Telescope\/} (Smith et al. 1997), strongly support that the origin of the
blue lobes could be an accreted population formed from the cluster cooling
flow or from episodes of star formation induced by accretion of stripped
material from several gaseous dwarf galaxies.

In this work we address the possible connection between star formation and
the cooling flow phenomenon through the study of the stellar populations in
CCGs. For this purpose we analyse line-strength gradients in Mg$_2$ (Faber et
al. 1985) and the \mbox{$\lambda$4000-\AA\ break} 
(D$_{4000}$, definition in Bruzual
1983) in a sample of 11 brightest cluster galaxies.  This paper is an
extension of a previous work presented in CGA95 after the inclusion of
extensive new spectroscopic data.

The galaxy sample, reduction of the data and error analysis are presented in
Sections~2 and~3. The nuclear line-strength and gradients are described in
Section~4. The discussion of these results is presented in Section~5, and our
conclusions are summarised in Section~6.
\mbox{$H_0 = 50$ km s$^{-1}$ Mpc$^{-1}$} has been assumed throughout the paper.

\section{Galaxy sample and observations}

We have obtained long-slit spectra of CCGs in two runs with different
telescopes. Since the aim of the observations was to get reliable
line-strength gradients, they were performed on dark nights. The first run
(August 1994) was carried out with the 3.5-m telescope at the German-Spanish
Astronomical Observatory at Calar Alto (Almer\'{\i}a, Spain).  We used 
the Cassegrain Twin Spectrograph with a blue coated TEK CCD in the red
channel. In the second run (December 1995) we observed with the 4.2-m WHT at
the Roque de los Muchachos Observatory (La Palma, Spain), using the blue arm
of ISIS with a blue coated TEK CCD.  A summary of the observations, including
relevant instrumental parameters, is given in
Table~1.

The sample of CCGs was chosen to cover a wide range of mass deposition rates
and complements our previous data (CGA95). It includes 8 galaxies in clusters
with cooling flows and 3 galaxies in clusters without cooling flows. A
description of the sample, including exposure times and position angles of
the spectrograph slit, is listed in Table~2. We have
re-observed the central galaxy of A~644 since previous measurements (CGA95)
were not clearly understood. We also observed G--K giant stars to guarantee
that our Mg$_2$ indices are in the appropriate photometric system (see
below).

\section{Data reduction and error analysis}

Reduction of the data was carried out using our own reduction package
\reduceme\ (Cardiel \& Gorgas 1998\footnote{See also:\\
http://www.ucm.es/OTROS/Astrof/reduceme/reduceme.html}). We followed a
standard reduction for long-slit spectroscopic data: bias and dark
subtraction, cosmic ray cleaning, flat-fielding, wavelength
calibration, C-distortion correction, sky subtraction, extinction
corrections, centring and binning of the spectra.  The process is
similar to that explained in CGA95, and its description is not repeated
here. However, we present a summary of the error sources and the
procedure followed for their estimation.

\subsection{Random errors}

We describe below the main sources of random errors.  Except for the inner
parts of the galaxies with long exposure times, these are strongly dominated
by photon statistics and read-out noise.  The contribution of additional
sources of random error is quite negligible.

(i) {\it photon statistics and read-out noise:\/} They have been
computed with the help of a new set of analytic formulae derived by
Cardiel et al.  (1998). Error images, which are created at an early
stage of the reduction procedure, are processed in parallel with data
images. The basic arithmetic manipulations performed on the data frames
are translated into the error frames following the usual error
propagation laws. The errors obtained using this method are in
excellent agreement with numerical simulations (see Cardiel et al. 1998
for details). In the outer parts of the galaxies the spectra were
spatially binned to guarantee maximum random errors
\mbox{$\Delta$D$_{4000} \la 0.15$} and \mbox{$\Delta$Mg$_2 \la 0.025$
mag}.

(ii) {\it flux calibration:\/} We observed four spectrophotometric standards
(Massey et al. 1988; Oke 1990) in each run. The derived calibration curves of
each run were averaged and we estimated the uncertainty in flux calibration
as the rms scatter among the indices measured with different curves
(\mbox{$\Delta$D$_{4000} \simeq 0.03$}, \mbox{$\Delta$Mg$_2 \simeq 0.004$}
mag for run~1, and \mbox{$\Delta$D$_{4000} \simeq 0.03$}, \mbox{$\Delta$Mg$_2
< 0.001$} mag for run~2).

(iii) {\it wavelength calibration:\/} Spectra were transformed to a
linear wavelength scale by using \mbox{$\sim 35$} arc lines fitted
by~5th order polynomials, with typical \mbox{rms~$\simeq 0.3$~\AA}
(run~1) and \mbox{rms~$\simeq 0.5$~\AA} (run~2). These scatters are
completely consistent with those obtained by cross-correlating the
central spectra of the single exposures of each galaxy.  These
uncertainties translate into \mbox{$\Delta$D$_{4000}<0.01$} and
\mbox{$\Delta$Mg$_2<0.001$} mag for both runs.

(iv) {\it radial velocities:\/} Radial velocities were computed by
cross-correlating the central spectra of the final co-added image of
each galaxy with different templates (usually \mbox{$\sim 4$} broadened
spectra of bright G--K giant stars). The different measurements were
obtained with a typical rms~\hbox{$\simeq 10$}~km/sec (run~1) and
rms~\hbox{$\simeq 20$}~km/sec (run~2), which has a negligible effect in
the measurement of the indices (\mbox{$\Delta$D$_{4000}<0.01$} and
\mbox{$\Delta$Mg$_2<0.001$} mag). For comparison, an error of
\hbox{$200$}~km/sec turns into \mbox{$\Delta$D$_{4000}\simeq 0.01$} and
\mbox{$\Delta$Mg$_2\simeq 0.004$} mag for both runs.

\subsection{Systematic errors}

Systematic errors in the measurement of spectral indices arise mainly
from the effects of spectral resolution (and velocity dispersion), flux
calibration, sky subtraction and contamination by nebular emission,
among others. Unfortunately, systematic errors do not allow the same
straightforward treatment used for the random errors.

(i) {\it spectral resolution and velocity dispersion:\/} The Mg$_2$
index, which is defined using relatively broad bandpasses, is quite
independent on spectral resolution and velocity dispersion corrections
(e.g.  Gorgas, Efstathiou \& Arag\'{o}n-Salamanca 1990; Gonz\'{a}lez
1993; Carollo, Danziger \& Buson 1993). The D$_{4000}$, due to its
large wavelength coverage, is completely insensitive to these effects.
Thus, these particular indices do not require any spectral resolution
correction.

(ii) {\it flux calibration:\/} In order to guarantee that our Mg$_2$
measurements are in a consistent photometric system, we observed a
sample of~39~(run~1) and~13~(run~2) template stars (typically G--K
giant stars) from the Lick library (Gorgas et al. 1993; Worthey et al.
1994). We compared our Mg$_2$ indices with those of the Lick group in
Fig.~1. We found a systematic offset of 0.021~mag (with
a rms scatter of 0.010~mag).  This value is almost the same when
analysing the data for both runs separately: \hbox{$0.020$ (rms
$=0.008$)} and \hbox{$0.021$ (rms $=0.021$)} mag for run~1 and~2,
respectively. This offset is similar to that reported in CGA95
(0.014~mag), and it is also consistent with those determined by other
authors:  e.g. 0.017~mag (Davies, Sadler \& Peletier 1993; Gonz\'{a}lez
1993), 0.010~mag (Carollo \& Danziger 1994). This
systematic offset is due to the fact that the Lick data were flux
calibrated using a tungsten lamp as a relative calibrator of the
instrumental spectral response and, thus, they are not in a true
photometric system. However, we have transformed our Mg$_2$ indices to
the Lick system to perform comparisons with Mg$_2$ published in the
literature.

Unfortunately, the Lick library does not cover the spectral region
corresponding to D$_{4000}$.  To check the reliability of the
photometric system employed in the determination of this spectral
feature we have compared the measurements of four stars in common with
CGA95. The scarcity of this sample only allow us to conclude that the
D$_{4000}$ indices measured here and those presented in CGA95 are,
within errors, in the same photometric system (offset~$=0.17$,
rms~$=0.27$).

(iii) {\it sky subtraction:\/} Since galaxy light levels are usually only a
few per cent of the sky signal in the outer parts of the galaxies, this
process constitutes one of the most important potential sources of systematic
errors when measuring line-strength gradients. This problem was
treated in CGA95, and we refer the interested reader to that
paper.  Although there is not a simple recipe to detect a systematic effect
in the sky subtraction, anomalous deviations of the line-strength indices in
the outer parts of the galaxies are indicative of its presence.  We have
corrected from the contribution of the galaxy light in the regions where the
sky levels where measured by fitting de Vaucouleurs laws to the galaxy
profiles. When necessary, we have subtracted from the sky spectra a scaled
and averaged galaxy spectrum. Even in the outermost regions of the galaxies
the effect of such corrections on the measured indices is nevertheless quite
small (\mbox{$\sim10$--$20$ per cent} of the random error).

(iv) {\it contamination by nebular emission:\/} Emission lines
present in the inner parts of some CCGs (e.g. [Ne\,{\sc
iii}]~$\lambda$3869, [S\,{\sc ii}]~$\lambda\lambda$\mbox{4069, 4076},
H$\delta$, [O\,{\sc iii}]~$\lambda$4959, [N\,{\sc i}]~$\lambda$5199)
must be removed from the bandpasses employed in the measurement of Mg$_2$ and
D$_{4000}$. For this purpose we have followed an interpolation procedure
based on a simultaneous fitting of a scaled template spectrum, a low order
polynomial and a Gaussian, in the region close to the emission lines (see
Fig.~2). The pixels affected by the presence of emission
where carefully replaced by the result of that fit (i.e.
\mbox{polynomial$+$template}). In some objects these corrections are not
negligible: e.g.  A~2597, \mbox{$\delta$D$_{4000}=0.06$},
\mbox{$\delta$Mg$_2=0.100$}; A~478, \mbox{$\delta$D$_{4000}=0.02$},
\mbox{$\delta$Mg$_2=0.040$}; Hydra~A, \mbox{$\delta$D$_{4000}=0.01$},
\mbox{$\delta$Mg$_2=0.010$} (\mbox{$\delta{\it Index}={\it Index}_{\rm
nel}-{\it Index}_{\rm el}$}, where \mbox{${\it Index}_{\rm el}$} is the index
measured without removing the emission lines, and \mbox{${\it Index}_{\rm
nel}$} the index obtained when the emission has been interpolated). This
procedure has been applied to the whole sample of galaxies
with emission lines in their central regions.

\section{Results}

\subsection{Measurement of the central indices}

Following the same procedure than in CGA95, and in order to avoid
aperture effects, we have obtained central D$_{4000}$ and Mg$_2$
indices employing a standard metric aperture size corresponding to
4~arcsec projected at the distance of the Coma cluster
($\simeq2.6\,$kpc). The central indices presented in CGA95 were not
corrected from emission lines. In order to get a fully consistent list
of central indices, we have revised the spectra of that sample which
exhibited emission lines. We have found that the effect is non
negligible in two objects, namely A~1795 (with
\mbox{$\delta$D$_{4000}=0.02$} and \mbox{$\delta$Mg$_2=0.026$~mag}, and
A~2199 (with \mbox{$\delta$D$_{4000}=0.01$} and
\mbox{$\delta$Mg$_2=0.009$}~mag).  In addition, we have transformed all
the Mg$_2$ measurements to the Lick system by applying the derived
systematic offsets.  A full list with all the available central
measurements and associated random errors, including the revised data
from CGA95, is given in Table~3.

\subsection{Line-strength gradients}
\label{sectiongradients}

We have measured D$_{4000}$ and Mg$_2$ gradients for all the galaxies
of the sample.  In order to obtain symmetric brightness profiles at
both sides of the galaxy centre, the spectra were shifted by a fraction
of pixel in the spatial direction. This process was not carried out with
the central galaxy of A~2626 since in this object a secondary nucleus
was also observed inside the slit at only $\sim 3.4$~arcsec from the
centre of the CCG.  In this case a simultaneous fit of two Cauchy
functions plus a low order polynomial allows us to determine the
location of the centre of the galaxy and the secondary nucleus, together
with the relative light contribution of each component nearby the
central region of the galaxy.  The spectra were spatially binned in the outer
parts of the galaxies to guarantee a minimum value of the
signal-to-noise ratio.

D$_{4000}$ and Mg$_2$ gradients are plotted in Figs.~3
and~4, respectively, where the galaxies are displayed in
decreasing mass deposition rate (label in the upper right corner of
each panel). Galaxies from CGA95 with reliable gradients are also
included in these diagrams. Filled circles and stars correspond to
measurements in different sides of the galaxies. Open circles refer to
secondary nuclei. The thick horizontal solid lines in the upper left
corner of the panels indicate the spatial extent of the emission lines,
when present. We define the emission
line region as the area where the emission lines were detected above
1~$\sigma$. The cut-off of this region is very sharp, so the
actual deffinition of its edge is not critical.

In most cases, indices measured at both sides of the galaxies agree
within the errors.  Error-weighted least-squares fits to all the
points, excluding data within the central 1.5~arcsec (affected by
seeing) and secondary nuclei, are shown as thin lines for all the
galaxies. In addition, we have computed {\it inner\/} and {\it outer\/}
gradients (i.e. {\it inside\/} and {\it outside\/} the emission line
region, respectively), in the galaxies with central emission lines.
These inner and outer gradients, plotted with a thick line, have been
forced to join at the radius where emission lines end. The results
obtained in all these fits, together with their formal errors, are
given in Table~4.

\subsection{Comparison with previous work}
\label{secpreviouswork}

We have carried out a comparison of our central D$_{4000}$ measurements
in CCGs (see Table~3) with the available data
in the literature. In Fig.~5 we plot D$_{4000}$
data from JFN87 and MO89, and \mbox{(U$-$b)}$_{\rm nuc}$ from MO92
against our central D$_{4000}$ indices.  There is a clear systematic
offset of $0.24$~(rms~$=0.04$) in D$_{4000}$ between the indices of
JFN87 and this work. Although error bars are not available for the
JFN87 data, assuming a typical random error of $\Delta {\rm D}_{4000}
\sim 0.04$, the derived offset is probably inconsistent with the
intrinsic errors and must be considered as a systematic effect.  On the
contrary, the D$_{4000}$ values from MO89 are consistent with our
measurements since the computed offset is $0.06 \pm 0.08$. The nuclear
\mbox{(U$-$b)} colours from MO92 display a very good correlation with
the D$_{4000}$.  A similar correlation was previously presented by
MO89.  This correlation can be easily understood since the
\mbox{(U$-$b)} colours measure essentially the same spectral break than
D$_{4000}$.  The dashed line is panel~5b is the
least-squares fit to the data:  \mbox{(U$-$b)$_{\rm nuc} = -1.03 (\pm
0.14) + 0.81 (\pm 0.07) {\rm D}_{4000}$}.  This linear relation will be
employed later in this paper to estimate central D$_{4000}$ from
\mbox{(U$-$b)$_{\rm nuc}$} colours.

Line strength gradients of Mg in brightest cluster galaxies have been
presented by Fisher, Franx \& Illingworth (1995, hereafter FFI95).  Our full
sample includes four galaxies in common with theirs, namely N~2832 (A~779),
N~6166 (A~2199), N~7720 (A~2634, ---note: this object was mislabelled as
N~7728 in CGA95) and the central cluster galaxy of A~496. In order to
compare their measurements of Mgb with ours of Mg$_2$ we have transformed Mgb
into Mg$_2$.  The atomic Mgb index is very well correlated with the molecular
Mg$_2$ index in early-type systems: Gonz\'{a}lez \& Gorgas (1995) report
${\rm Mg}_2{\rm (mag)} \simeq 0.066 \, {\rm Mgb(EW)}$.  The comparison with
FFI95 data is shown in Fig.~6. The measurements for the central
galaxy in A~779 exhibit an excellent agreement. However, this is not the case
for the central galaxies of A~496, A~2199 and A~2634.

The discrepancies in A~496 and A~2199 are readily understood in terms of
contamination by the emission line [N\,{\sc i}]~$\lambda$5199. Unfortunately,
this line is centred in the red continuum bandpass of the Mgb index, and its 
presence translates into an
overestimation of the index value due to an enhancement of
the measured continuum level. The effects of this emission have been
studied in detail by Goudfrooij \& Emsellem (1996). Since in
Fig.~6 we have derived Mg$_2$ from the Mgb indices of
FFI95, these Mg$_2$ values are also overestimated.  Although we cannot
correct the FFI95 indices from the effect of the emission line, we can
reproduce their measurements by measuring Mgb in our spectra without
removing the contamination of [N\,{\sc i}]~$\lambda$5199, and then
transforming these Mgb values into Mg2 indices. The effect of such
procedure is shown in panels~6b and~6c as
dotted arrows.  It is clear from the figures that in this case the
agreement between FFI95 data and our gradients is then fairly good for
A~2199 and A~496. In fact our indices lie between those determined by
FFI95 along the major and minor axis.

It is important to note that the effect of [N\,{\sc i}]~$\lambda$5199 in
Mg$_2$ is much smaller than in Mgb.  Goudfrooij \& Emsellem (1996) showed
that a strong EW of [N\,{\sc i}]~$5199 \sim 1.8$~\AA\ translates into
$\Delta$Mgb$ \; \sim 2$ but only $\Delta$Mg$_2 \sim 0.03$ (transforming this
$\Delta$Mgb into $\Delta$Mg$_2$ using the linear relation described above
gives $\Delta$Mg$_2 \sim 0.13$, which is four times larger).

Nevertheless, the Mg$_2$ gradient for the central galaxy of A~2634 does
not show the same agreement found in the previous objects (note that, in
this case, Mg$_2$ is not affected by the emission lines). Our
measurements are clearly below the FFI95 data. Additional central
Mg$_2$ measurements for A~779, A~2199 and A~2634 are also available
from Trager (1997) ---plotted as a short horizontal full line--- and
Lucey et al. (1997) ---short horizontal dashed line. These central
indices agree with our central values for A~779 and A~2199, confirming
the previous comparison with the FFI95 data. In addition, the central
Mg$_2$ values of A~2634 from Trager (1997) and Lucey et al. (1997) lie
in the region between the FFI95 data and ours.  This discrepancy, which
affects mainly the absolute values of the Mg$_2$ index, but not the
shape of the radial gradient, is not understood, so the caution is
needed when interpreting the A~2634 data.

\section{Discussion}

\subsection{Central spectral indices}
\label{sectiondiscussionindex}

Different authors have reported correlations of central blue colours and
spectral features with mass deposition rates: JFN87 found a correlation
between D$_{4000}$ and the mass flow rate ($\dot{M}_{\rm V}$) within their
spectrograph slit; MO89, MO92 and McNamara (1997) shown that central
$U-B$ colour excesses in CCGs immersed in clusters with high mass
deposition rates also correlate with $\dot{M}$; in CGA95 we confirmed the
correlation with D$_{4000}$ and presented a correlation with Mg$_2$.

In Fig.~7 we have plotted all the central measurements
listed in Table~3, as a function of the X-ray derived
mass deposition rate. In panels~7a and~7b
we have used the total $\dot{M}$ values, whereas in
panels~7c and~7d we computed the central
mass deposition rates in the same fixed metric aperture size employed in the
measurement of the central indices, assuming $\dot{M} \propto r$.  We have
also included in panels~7a and~7c central
D$_{4000}$ estimates of eight additional galaxies obtained from the nuclear
\mbox{(U$-$b)} colours of MO92. For this purpose we have used the linear fit
derived in Section~\ref{secpreviouswork}. These new indices are listed in
Table~5.  Filled and open symbols in
Fig.~7 correspond to CCGs with and without central emission
lines, respectively.

Although, as mentioned above, central
D$_{4000}$ and Mg$_2$ indices are well correlated with mass deposition rate
in the sense that both indices decrease when the mass flow increases, it is
interesting to note that some objects lie off the main relation. In
particular the cooling flow galaxies without emission lines A~644, A~2029 and
A~2142 are found above the relation, being their central indices fully
consistent with those of central galaxies in clusters without cooling flow.
Simultaneously, there are also two cooling flow galaxies with emission lines,
Hydra~A and N~1275, which also seem to depart from the relation with
$\dot{M}$.

To perform a simultaneous study of the variations in D$_{4000}$ and Mg$_2$,
we present an index--index diagram in Fig.~8.  In
panel~8a we plot the central indices with their errors for
the whole sample.  In the following panels of this figure we have suppressed
the galaxy names and error bars for clarity.  Since the CCGs exhibit a clear
sequence in this index--index plane, in panel~8b we compare
this sequence with that displayed by early-type galaxies. The small dots
correspond to a sample of 135 elliptical and S0 galaxies observed by Kimble,
Davidsen and Sandage (1989; note: we have transformed their Mgb data into
Mg$_2$ using the same relation employed in section~\ref{secpreviouswork}).
As we already shown in CGA95, CCGs in cooling
flow clusters populate the D$_{4000}$--Mg$_2$ plane following a different
sequence to that displayed by elliptical galaxies.  The enlargement of the
galaxy sample allow us to gain new insights into this study. In
particular, it is clear from the figure that the CCGs in cluster without
cooling flows (open circles) lie in the same region than the giant
ellipticals (high Mg$_2$ and D$_{4000}$ values).  The central galaxies in
clusters with cooling flows which do not exhibit emission lines in their
central parts (open stars, with symbol size proportional to mass deposition
rate) are also compatible with the location of the non cooling flow galaxies,
even for high $\dot{M}$, like A~2142 and A~2029. On the other hand, cooling
flow galaxies with central emission lines (filled circles, with symbol size
proportional to $\dot{M}$) depart from this location, being the largest
accretors the galaxies which show the largest deviations.  Using the same
symbols, we plot in panel~8c the D$_{4000}$ obtained
from \mbox{(U$-$b)$_{\rm nuc}$} colours of MO92 (as explained above).
It is interesting to note that some cooling flow galaxies with only
moderately high $\dot{M}$ (Hydra~A and N~1275) display quite low
D$_{4000}$ values.  Thus, it is clear that cooling flow galaxies with
emission lines do not represent a simple one-parameter sequence in mass
deposition rate.

We investigate now the behaviour of the spectral indices as a function
of age and metallicity by studying the evolution of single bursts of
star formation. 
We have used the galaxy isochrone
synthesis spectral evolution library, GISSEL96, from Bruzual \& Charlot
(1997). In panel~8d we plot the evolution of three
single bursts of star formation (full lines) with different metallicities
([Fe/H]=$-$0.64, $-$0.33 and 0.09).  The model sequences 
are labelled with small
circles which indicate the age of the burst in Gyr. Mg$_2$ values in the
models are computed using the empirical fitting functions derived by Gorgas
et al. (1993) and Worthey et al. (1994).  Unfortunately, this type of fitting
functions are not available for the D$_{4000}$, and a non-negligible
systematic offset between the predictions of the models and the data could be
present.  For simplicity, we represent the sequence of normal
ellipticals and S0 of Kimble et al.  (1989) by a dashed parallelogram. Taking
into account the models, we have drawn two vectors to indicate changes 
of age and metallicity in this D$_{4000}$--Mg$_2$ plane and, interestingly,
both parameters are not completely degenerated in this diagram. It suggests
that normal ellipticals represent a metallicity sequence, whereas
the cooling flow objects describe a locus quite parallel to the age
evolution. This strongly suggests that the observed spectral properties
of the cooling flow galaxies can be understood as a consequence of star
formation.

Following the procedure described in Gorgas et al. (1990) and in CGA95,
we have modelled the expected changes in the central D$_{4000}$ and
Mg$_2$ indices when a normal CCG, initially without cooling
flow, undergoes star formation as a result of the cooling flow
---panel~8e. 
All the models describe the effect of such star formation overimposed on a
galaxy with D$_{4000}$ and Mg$_2$ indices equal to the averaged values
obtained with our subsample of five CCGs in clusters
without cooling flows (i.e., typical of old stellar populations).  
Dotted lines represent the effect of a continuous
star formation during the last 0.1~Gyr, 1.0~Gyr and 10.0~Gyr (from bottom to
top). Full lines correspond to single bursts occurred 0.01~Gyr, 0.1~Gyr and
1.0 Gyr ago (also from bottom to top). In all these models a Scalo (1986) IMF
and solar metallicity were adopted.  Finally, the dashed line indicates the
effect of a power-law spectrum ($F_{\lambda} \propto \lambda^{\alpha-2}$,
with $\alpha=0.5$). No intrinsic reddening has been taken into account in the
models.  Small dots in the model lines indicate the fixed fraction $f_V$
(given by the numbers) of the $V$~light that comes from the 
accretion population (i.e., the newly-formed stars). 
In the case of a continuous star formation, this fraction $f_V$ can
be related to the SFR using the expression
\begin{equation}
\label{eqcontinuous}
{\it SFR} = \frac{f_V \, L_V \, (M/L)_{\rm AP}}{t},
\end{equation}
where $L_V$ is the total V~luminosity of the galaxy, $(M/L)_{\rm AP}$ is the
stellar mass-to-light ratio of the accretion population, given by the models,
and $t$ is the time over which the star formation has taken place. Similarly,
the parametrization for the models involving a single burst of
star formation is given by
\begin{equation}
\label{eqburst}
{\it M}_{\rm AP} = f_V \, L_V \, (M/L)_{\rm AP},
\end{equation}
where ${\it M}_{\rm AP}$ is the total mass in stars formed in the accretion 
population.

Panel~8e reveals that the sequence of cooling flow
galaxies does not seem to be very well reproduced with a continuous
star formation during the last 1--10~Gyr. Either a {\it young\/} ($<
1.0$~Gyr) continuous star formation, or simply a recent burst of star
formation $\sim 0.1$~Gyr ago, can match the observed central indices.
In the lower right corner of panel~8e we have also
plotted the effect of intrinsic reddening in the measurements. As an
example, we represent the corrections for an hypothetical object with
D$_{4000}=1.7$ and Mg$_2 = 0.300$ as a function of $E(B-V)$ (the colour
excess ranges from 0.0 to 1.0, with ticks spaced 0.1 units). As it
could be expected, the effect on Mg$_2$ is almost negligible, whereas
the contrary is true for the D$_{4000}$. Although Hansen et al. (1995)
have reported \mbox{$E(B-V) \leq 0.1$} for Hydra~A, other authors have
found non-negligible reddening values ($E(B-V) \sim$~0.1--0.6) for
several cooling flow galaxies (e.g. Hu 1992; Allen et al.  1995;  Allen
1995; Donahue \& Voit 1993; Voit \& Donahue 1997). If we allow for the
existence of internal reddening, the case for recent bursts instead of
a continuous star formation is even stronger.  Nevertheless, it is
important to keep in mind that the effect of reddening in the stellar
continuum is usually lower than in the emission lines, e.g.  Calzetti
(1997) reports $E(B-V)_{\rm stellar} = 0.44 \, E(B-V)_{\rm gas}$ in
typical star forming regions.

However, we need to be cautious since, apart from intrinsic reddening,
there are additional sources of uncertainty present in the models.
First, there is some uncertainty in the starting point of the model
sequences, since the dispersion in the central indices of CCGs without
cooling flows is probably real.  Second, we have already mentioned that
the modelling of D$_{4000}$ is somewhat uncertain given the lack of
empirical calibrations, thus some systematic effects could be present.
And finally, the models assume solar metallicity {\it and\/} solar
element ratios: if there is an enhancement of [Mg/Fe] in the cooling
flow material, the predicted Mg$_2$ indices could be underestimated.
For these reasons, it seems premature to rule out competely, from the
models shown in panel~8e alone, a continuous star
formation (lasting some Gyrs) as the source of the blue light.

Considering the relatively short wavelength baseline between the D$_{4000}$
and Mg$_2$ indices, the contribution of a power-law spectrum could also
reproduce the observed spectral changes. This result is not surprising since,
in a short wavelength interval, the spectrum of an early-type star can be
approximated quite well by a power law (see also Crawford \& Fabian 1993).
However, as we have already discussed in the introduction, the absence of
polarization in the blue light of A~1795 (McNamara et al. 1996b), does not
give strong support to this alternative. In addition, some central galaxies
show clear signatures of strong Balmer absorption lines (e.g. N~1275,
Crawford \& Fabian 1993; Hydra~A, Hansen et al. 1995, Melnick et al.
1997).  Allen (1995) found that the strong UV/blue continua in
the spectra of a small sample of CCGs were better described by young stars
than by power-law emission models. This author also reported positive
detection of Wolf-Rayet features in two CCGs (those in
A~1068 and A~1835), which favour the idea of the formation of massive stars.
In this sense, Voit \& Donahue (1997) have measured a weak He\,{\sc
ii}~$\lambda$4686 emission in A~2597. We also find evidences of such helium
feature in the central spectrum of the dominant galaxies in A~1795 and in
A~2597 (see Fig.~9)

The role of metallicity and the choice of IMF on the model predictions
plotted in panel~8e is explored in
panel~8f. We show the GISSEL96 (Bruzual \& Charlot
1997) predictions for different metallicities (as explained in the
panel legend), and two IMFs, namely Salpeter (1955) ---thin lines---
and Scalo (1986) ---thick lines---, for the single burst model with an age of
0.1~Gyr. The predictions of the models do not depend strongly on
the choice of both [Fe/H] or the IMF.

In order to obtain further constraints on the range of acceptable
models, we need to explore a larger wavelength range. We have done so
using UV and near-IR data from the literature (see
Fig.~10 and description in the figure caption).
Although a continuous star formation lasting 10~Gyr could explain the
observed optical and near-IR observations, the UV~fluxes are
underpredicted.  Recent bursts with ages $\sim 0.1$--$0.2$~Gyr are
simultaneously compatible with both the UV and IR fluxes, and with the
measured central line-strength indices. Note however, that at UV
wavelengths the effects of dust extinction can be quite dramatic even
for moderate colour excesses:  using the interstellar reddening curve
of Savage \& Mathis (1979), and assuming $E(B-V)=0.2$ and $E(B-V)=0.4$
we estimate that the measured UV fluxes need to be corrected upwards by
factors of $\sim 5$--6 and $\sim 30$--20, respectively. In these cases,
is quite obvious that the UV continuum would constrain the upper limit
of the burst age to be $< 0.1$~Gyr. This result agrees with the
behaviour of D$_{4000}$ and Mg$_2$ with $E(B-V)$ in
Fig~8e, indicating that a recent burst would be
strongly favoured against constant star formation.

Another piece of evidence which can help to understand the behaviour of
the spectral indices of the CCGs comes from the emission line
properties.  
Using line-ratio diagrams HBvBM
classified the emission nebulae at the centres of central dominant galaxies
into two types. In particular, Type~I corresponds to objects with relatively
high [N\,{\sc ii}]$\lambda$6584/H$\alpha$ ($\simeq 2.0$) and 
[S\,{\sc ii}]$\lambda$6717/H$\alpha$ ($\simeq 0.7$), whereas Type~II objects
exhibit lower line-ratios ([N\,{\sc ii}]$\lambda$6584/H$\alpha$ $\simeq 0.9$
and [S\,{\sc ii}]$\lambda$6717/H$\alpha$ $\simeq 0.4$).
In
panels~8g and~8h we have plotted the
available HBvBM types for the galaxies in our sample and in MO92 respectively.
There is a clear dichotomy in the way Type I (plotted with ``1'' symbols) and
Type II (plotted with ``2'' symbols) galaxies populate the central cluster
galaxy sequence. This result is not surprising since Type I objects are
associated to clusters with low X-ray and H$\alpha$ luminosities, small
optical nebulae and low mass deposition rate, whereas the contrary is true
for Type II objects (HBvBM, Baum 1992, Donahue \& Voit 1997). 

Another clue comes from the morphology of the blue excess light. 
Recently, McNamara (1997) has established a morphological classification based
on the study of the surface brightness distributions and colour structure in
cooling flow CCGs. This classification consists in four
morphological types ordered by increasing geometrical complexity: (1) point
source (e.g. A~2199, A~2052) with unresolved blue nuclei and nebular line
emission, low nebular luminosity, with modest blue anomalies;  (2) disk
(e.g.  Hydra~A) unusual in cooling flow galaxies and characterised by a disk
of gas and young stars rotating around the nucleus; (3) lobe (e.g.  A~1795,
A~2597) with bright, blue lobes of optical continuum separated several
kiloparsecs from the nucleus; and (4) irregular-amorphous (e.g. N~1275 may be
in transition between types~3 and~4). In Fig.~8g we have
also plotted the four morphological types in the neighbourhood of the galaxies
which display such characteristics.  Taking into account the D$_{4000}$
exhibited by N~1275, we have tentatively located Type~4 near PKS$0745-191$.
Very interestingly, McNamara (1997) has speculated about the possibility
that the structural types reflect actually an evolutionary sequence. We will
discuss this point in more detail in
section~\ref{section:wholepicture}.

\subsection{Line-strength gradients}
\label{sectiondiscussiongradients}

Although line-strength gradients in Mg$_2$ (e.g. Faber 1977; Gorgas et al.
1990; Gonz\'{a}lez 1993; Davies et al. 1993; Carollo et al. 1993; FFI95;
Gonz\'{a}lez \& Gorgas 1995) and in D$_{4000}$ (e.g. Munn 1992; Davidge \&
Grinder 1995) are usually found to be linear with $\log(r)$ in elliptical
galaxies, in CGA95 we showed that some cooling flow galaxies exhibit a clear
slope change at intermediate radii. With the inclusion of the new data
presented in this paper (see Figs.~3 and~4),
we find that this change of slope is clearly related to the presence of
emission lines in the central regions of CCGs.

We measured the radius at which the change of slope takes place using
error-weighted least-squares fits to two straight lines forced to join
at a variable radius. The method gives the {\it break radius\/} which
minimises the residual variance of the fits.  We estimated the
uncertainty in the break radius using Monte Carlo simulations that take
into account the measurement errors in the indices (typically $\simeq
1000$ simulations per gradient). For the emission line galaxies with
reasonably good data beyond the emission region, we have compared in
Fig.~11 the break radius with the observed
emission-line radius (as defined in section~4.2).   The
figure reveals that the break radii are almost coincident with the
emission-line radii, with the former being slightly larger. 

It is quite evident from Figs.~3 and~4
(see also the fitted gradients tabulated in
Table~4), that cooling flow galaxies {\it with
emission lines} exhibit flat and even positive gradients in the region
where the emission is detected (e.g. A~478, A~1795, A~2597, Hydra~A,
A~85, A~2199). In fact, these inner gradients (measured in the emission
line region) seem to be correlated with the mass deposition rate, i.e.
more positive with increasing $\dot{M}$ (see Figs.~12a
and~12b). In the outer parts of these galaxies, where
emission is not observed, the derived mean D$_{4000}$ gradient is
consistent with that observed in normal elliptical galaxies (see
Fig.~12c). However, the mean Mg$_2$ gradient in the
outer regions of cooling flow galaxies with emission lines seems to be
marginally stepper than the mean gradient for normal ellipticals
(Fig.~12d), although the quality of the Mg$_2$ data at
large galactocentric distances is not good enough to reach a firm
conclusion.  A couple of galaxies (A~1126 and A~262) seem to have
slightly stepper gradients in the inner parts than in the outer
regions, although this result is not very significant due to the poor
S/N in the outer spectra.  In addition, mean gradients in central
dominant galaxies with cooling flow but {\it without emission lines\/}
(e.g.  A~644, A~2142) are similar to those in galaxies without cooling
flows and in elliptical galaxies.

If we assume that the observed slope changes in the line-strength
gradients are due to radial index variations induced by star formation
in the emission-line region, we can use that information, coupled with
the star formation models, to estimate the amount and spatial profile
of the accretion population. We assume that the gradients determined in
the outer regions represent the {\it underlying\/} gradient, i.e., that
of the galaxy before the cooling-flow induced star formation takes
place. That is justified by the similarity of these outer gradients
with those of ``quiescent'' galaxies (i.e., both normal giant ellipticals and
CCGs without cooling flows).  In this analysis we have exclusively
concentrated on the D$_{4000}$ gradients, which  are much better
determined than the Mg$_2$ ones. 

As discussed in section~5.1, a single burst model with an age $\simeq
0.1\,$Gyr, solar metallicity and Scalo IMF is able to reproduce the
observed spectral properties of the cooling flow galaxy sequence. We
have employed this particular model to tentatively obtain the spatial
profile of the accretion population.  First we computed the averaged
indices inside the emission line region (simulating a circular aperture
by weighting with light and radius), both from the fitted inner
gradient and from the extrapolation of the outer gradient into the
emission line region.  Then, we used the star formation model to derive
the mean $f_V$ of the accretion population required to reproduce the
observed changes in the indices (i.e., the difference between the
measured values and those estimated from the extrapolation of the outer
gradient). Assuming that the absolute magnitude corresponding to the
old stellar population in the emission line region  
is $\langle M_V \rangle \sim
-22.3$~mag\footnote{Hoessel, Gunn \& Thuan (1980) reported that the mean
magnitude of a sample of brightest cluster galaxies inside a metric radius of
$16/h_{60}$~kpc is $\langle M_V \rangle = -22.68 \pm 0.03$~mag. In addition,
Hoessel (1980) showed that the internal regions of brightest cluster galaxies
are adequately described by a modified Hubble law of the form
$I(r)=I_c/(1+r^2/r_c^2)$, where $I_c$ is the central intensity, $r_c$ the
core radius and $r$ the angular distance from the centre. Assuming this
intensity profile, the integrated luminosity inside a radius $r$ is then
given by $L(r)=\pi \, I_c \, r_c^2 \, \ln(1+r^2/r_c^2)$. Introducing the
value of the Hubble constant adopted in this paper (\mbox{$H=50 \; {\rm km}
\;  {\rm s}^{-1} \; {\rm Mpc}^{-1}$}), and the mean core radius derived by
Hoessel (1980), $\langle r_c \rangle = 2.23/h_{60}$~kpc, we obtain
\mbox{$\langle M_V \rangle \sim -22.3$~mag} inside a radius \mbox{$r\sim
11/h_{50}$~kpc}, which is the average radius of the emission line regions
in our sample.}, the mean $f_V$ value is then employed to compute the
total $V$~luminosity ($L_V$) of the galaxy plus the burst. Using
Eq.~\ref{eqburst}, and the mass-to-light ratio of the accretion
population given by the model, we obtain the total mass in stars of the
accretion population formed in the last burst. Note that, since the
excess blue light is completely dominated by the youngest stars, this
method is only sensitive to the stars formed in the most recent burst.

In Fig.~13 we have plotted the estimated total amount
of mass transformed into stars in the emission-line region versus the
mass deposition rate computed in the same area (assuming that
\mbox{$\dot{M} \propto r$}). There is an increase in the mass
transformed into stars with $\dot{M}$, with Hydra~A lying somewhat
outside this trend.  Interestingly, the total mass accreted by the
cooling flow in a $\sim0.1\,$Gyr period inside the emission line region
is remarkably similar to the mass in new stars 
(see section~\ref{section:wholepicture}).

Assuming spherical geometry, it is possible to apply the same technique
to obtain the amount of mass transformed into stars in concentric
shells at different radii. Using a deprojection algorithm described in
Appendix~A, we have derived the spatial profiles for the density of
mass transformed into stars (Fig.~14).  We
also show (dashed line) the expected density profile of the mass
deposition assuming $\dot{M}(<r) \propto r$.  Interestingly, the
derived profiles seem to be quite parallel to the dashed line (the mean
slope for all the galaxies excluding Hydra~A is $-$2.08,~\mbox{${\rm
rms} = 0.18$}), whereas Hydra~A exhibits a statistically significant
steeper trend.  This result strongly suggests a direct connection
between the mass deposition and the star formation profiles.
Indeed, 
figures~13 and~14 suggest a
very close link between the star formation and the cooling flow
phenomenon. Such close connection would be difficult to understand
without most of the gas accreted in the emission-line region being
transformed into stars. 

\section{Summary: towards a complete picture?}
\label{section:wholepicture}

Many pieces of the cooling flow puzzle are now available.  We summarise
some of them here before trying to sort out a scenario able to
explain the observational data.

(i) Central D$_{4000}$ and Mg$_2$ indices in cooling flow galaxies are
correlated with mass deposition rate when emission lines are found in
the central regions of such galaxies.

(ii) The nuclear indices of the cooling flow galaxies in our sample
which do not exhibit emission lines do not follow this correlation with
$\dot{M}$.  The central line-strengths of both CCGs in clusters without
cooling flows and cooling flow galaxies without emission lines are
consistent with the values observed in giant ellipticals.

(iii) The central D$_{4000}$ and Mg$_2$ indices of elliptical galaxies
define a relatively narrow trend in an index-index plane. Cooling flow
galaxies with emission lines exhibit a clear sequence in this diagram
which departs from the locus defined by elliptical galaxies. When
interpreted using stellar population models, this provides strong
evidence for star formation in these galaxies.

(iv) Although line-strength gradients are usually found to be linear with
$\log (r)$ in elliptical galaxies, D$_{4000}$ and Mg$_2$ gradients in cooling
flow galaxies with emission lines show a clear slope change in the region
where the emission is detected. In fact, these inner gradients seem to be
correlated with the mass deposition rate. In the outer parts of these
objects, where emission is not observed, the derived mean line-strength
gradients are consistent with those in elliptical galaxies.

(v) Mean gradients in cooling flow galaxies without emission lines are
similar to those in CCGs without cooling flow and those in giant
elliptical galaxies.

(vi) The presence of emission lines must be in some way related to the
cooling flow phenomenon since H$\beta$ (JFN87) and H$\alpha$
luminosities (HBvBM; Donahue \& Voit 1997) are correlated with
mass deposition rate.  However, considering the high scatter in line
luminosities for a given $\dot{M}$, the relationship does not seem to be a
simple one (Baum 1992). 

(vii) Whichever mechanism is responsible for the dilution of the
measured spectral indices, the correlations of the central indices and
gradients with the mass deposition rate in the emission-line region add
further weight to the idea of a link between the blue excess and the
cooling flow.  Nevertheless, as Crawford \& Fabian (1993) pointed out,
the fact that there are central galaxies in clusters with high
$\dot{M}$ but without blue excess suggests that the cooling flow alone
cannot be responsible for the blue light. In particular, these authors
noted the existence of strong cooling flow galaxies, with (e.g. A~478)
and without (e.g. A~2029) emission-line nebulae, which apparently did
not exhibit a blue continuum excess. However, our D$_{4000}$ and Mg$_2$
gradients in the central galaxy of A~478 does exhibit a clear positive
behaviour (i.e.  blue excess) in the emission-line region. An important
question is whether there are any cooling flow galaxies with
emission-line nebulae which do not exhibit a blue excess, and vice
versa. So far this does not seem to be the case.

(viii) Through a detailed analysis of the emission-line nebula in the central
cluster galaxy of A~2597, Voit \& Donahue (1997) conclude that hot stars are
the most likely ionizing source of the gas if some mechanical form of heating
(flowing from the hot intracluster medium into the cooler nebula) supplements
photoelectric heating.

(ix) With the available data, there is a clear dichotomy in the way
cooling flow galaxies populate the D$_{4000}$--Mg$_2$ plane depending on
their emission-line properties as described by the classification
scheme of HBvBM. 
The relative position of Type~I and Type~II objects in 
Fig.~8g strongly suggests that recent
star formation (and therefore photoionization by hot stars) is
the key to understand the observed dichotomy.

(x) The excess of blue light (i.e. line-strength dilution) in the continuum
spectra of CCGs is consistent with recent star formation
episodes (e.g. Bertola et al. 1986; JFN87; Shields \& Filippenko 1990;
Crawford \& Fabian 1993; Hansen et al. 1995; Allen 1995; Melnick et al. 1997;
Figs.~8e and~10 ---this work).

(xi) If the observed index variations are due to star formation in the
emission-line region, the existence of correlation between the central
indices and the line-strength gradients with the mass deposition rates
implies that the cool gas must be transformed into stars with a similar
efficiency from galaxy to galaxy. In addition, the radial profiles of
the star formation density  (Fig.~14) are
remarkably parallel to the expected density of the mass deposition
profile. Both results are difficult to understand unless a large
fraction ({\it probably most\/}) of the cooling flow gas accreted into
the emission-line region is converted into stars.

(xii) A large fraction (\mbox{60--70\%}) of cD galaxies in cooling flows are
radio-loud, whereas this probability decreases to 20\% for cD galaxies in
non-cooling flow clusters, and 14\% for typical elliptical galaxies in
clusters (Burns et al. 1997).  Since the kinetic energy of the radio plasma
can be a significant fraction of the thermal energy in the cooling flow, the
correlation between radio emission and the presence of cooling flows is
probably indicating the existence of interactions between the radio plasma and
the centres of such cooling flows. However it is important to note that there
is only a weak correlation between the 1.4~GHz radio power (HBvBM)
and the \mbox{6-cm} radio power (Burns 1990) with $\dot{M}$.

(xiii) The CCGs in A~1795 and A~2597 host blue lobes of optical
continuum along the edges of their radio lobes (McNamara \& O'Connell
1993; Sarazin et al. 1995; McNamara et al. 1996a, 1996b).  Through the
analysis of numerical simulations De Young (1995) has shown that the
blue continuum can be due to young stars produced by jet-induced star
formation.  This author has suggested that the most appealing scenario
consists in short-lived jets (with ages less than 10$^7$ yr), with the
high-pressure cooling flow being responsible for the disruption of the
jets.

(xiv) Using an X-ray colour/deprojection technique, Allen \& Fabian
(1997a, 1997b) have proved that all the cooling flow galaxies in their
sample exhibit significant central concentrations of cooling gas. These
authors find that intrinsic X-ray absorption increases with decreasing
radius, and that significant excess absorption is not present in non
cooling-flow clusters.  The estimated times required for the cooling
flows to accumulate the observed absorbing gas are typically of only a
few 10$^8$ years.

Summarizing, all these results suggest that very likely the presence of
cooling flows, emission lines, blue excesses, radio emission, soft
X-ray absorbing gas and star formation are different aspects of the
same phenomenon.  As we have previously mentioned
(section~\ref{sectiondiscussionindex}, Fig.~8g),
McNamara (1997) has suggested that his morphological classification of
the blue structure in cooling flow CCGs reflects an evolutionary
sequence. This author proposes the following scenario: central cluster
galaxies accrete $\sim 10^8 M_{\odot}$ of gas from a gas-rich dwarf
galaxy or a cooling parcel of hot gas. The accreted gas flows to the
galaxy centre where it triggers a radio source (by providing the
necessary fuel), which subsequently induces a burst of star formation
due to the rapid collapse of cold clouds compressed by shocks along the
expanding radio source (type~3, lobe). In a few $\sim 10^7$~yr the
lobes disperse (type~4, amorphous), leading finally to an unresolved
point source (type~1). The type~2 (disk) objects do not fall inside
this evolutionary sequence. As McNamara (1997) points out, Hydra~A is
an example of tidal accretion and likely not accretion from the cooling
flow. In this sense, Melnick et al.  (1997) have proven that the
nuclear emission lines in this galaxy originate in a gaseous disk-like
structure with a rotational velocity of $\sim 300 {\rm km} {\rm
s}^{-1}$.  Under this hypothesis, it is not surprising that the
behaviour of the central galaxy of Hydra~A in Figs.~7,
12, 13 and~14 is
different from that of the remaining cooling flow objects. The locus of
N~1275 in Figs.~7a and~7c is also
suspicious. However this object exhibits such unusual properties that
it has been interpreted either as a system of colliding galaxies (Baade
\& Minkowski 1954; Minkowski 1957; Rubin et al.  1977; Hu et al. 1983;
Holtzman et al. 1992), two superposed galaxies without interaction (De
Young, Roberts \& Saslaw 1973), or even an exploding galaxy (Burbidge,
Burbidge \& Sandage 1963; Burbidge \& Burbidge 1965). Thus, it is
likely that the morphology displayed by this galaxy is more complicated
than that naively expected from the cooling flow phenomenon.

If this scenario is true, an important question is why are we observing
so many objects with signatures of recent star formation.  Given the
relatively large fraction of type~3 objects observed in our sample (at
least two: A~1795 and A~2597), and considering the timescale for the
transition of a CCG throughout this stage (a few $\sim 10^7$~yr), the
scenario proposed by McNamara (1997) is not tenable unless radio
triggered star formation takes place several times during the lifetime of
the central galaxies. Some indication that this could well be the case
comes from the study carried out by Allen (1995) who found that the
number ratios of stars of different spectral types needed to reproduce
the UV/blue continua in a sample of CCGs also favour a scenario in
which star formation occurs in a series of bursts.  The analysis we
have presented here (section~5) also suggests that a young burst
($\sim$0.1~Gyr of age)  is able to explain the observed spectral
properties of the emission-line CCGs better than a more continuous star
formation. The population synthesis models, in conjunction with the
spectral data, also indicate that the total mass of cooled gas
transformed into stars is roughly equal to the mass accreted inside the
emission-line-region over a $\sim10^8$~yr period.  Interestingly, the
accumulation time-scales derived by Allen \& Fabian (1997a, 1997b) to
reproduce their calculated masses of absorbing gas is also only a few
$10^8$~yr.  Since the observed blue light would be completely dominated
by the stars formed in the last burst, it is possible that a sequence of
bursts has occured before the most recent one, fueled by the accreted
gas. Since cooling flows can be disrupted by cluster-cluster
interactions (Burns et al. 1997), it is also possible that the sequence
of bursts takes place only occasionally. As Crawford (1997) pointed out,
if rich clusters were built up hierarchically, the subsequent mergers
of the subclusters could have disrupted earlier cooling flows, likely
high-redshift radio sources (Bremer, Fabian \& Crawford 1997).
Therefore, strong cooling flows could have existed in today's rich
clusters only over the last $10^9$ years, allowing for a few bursts of
star formation (spaced by a few $10^8$ years) 
having taken place during this period.

This scenario accounts for the fate of the gas accreted
within the region where emission lines are observed (i.e., the inner
$\sim20\,$kpc): it accumulates over periods of a few $10^8$ years and
then most of it (perhaps all) forms stars in bursts. Outside this
region, star formation signatures are not observed, perhaps because the
radio source is unable to trigger star formation at large distances.
What happens to the gas depossited there is not clear:  some of it is
observed as cool material out to $\simeq100\,$kpc (Allen \& Fabian
1997a, 1997b), but our lack of knowledge of the physics of star formation
prevents us from understanding why (massive) stars are not formed. 

This scenario could also provide a tentative explanation of why there
are CCGs in clusters with strong cooling flows that do not show any
signs of recent star formation: a powerful radio source is also
necessary to trigger or catalise the formation of (massive) stars.  The
cooling flow provides the fuel, but without a central engine the radio
source, needed to kick-start the star formation, would not develop. The
poor correlation between radio power and star formation indicators
(e.g., H$\alpha$ fluxes) could be accounted for by the difference in
time-scales involved in the radio phenomenon ($\sim10^7$ yrs) and the
star-formation signatures ($\sim10^8$ yrs).  There could also be some
delay between the onset of a powerful radio source and the begining of
the star formation, which could explain the few cases where a radio
source is present in cooling galaxies without clear signs of excess
blue light.

We must stress that this picture, although very suggestive, is highly
speculative, and further observational work is clearly needed.  
Our study of optical absorption features has proven to be
an excellent tool to investigate the effect of cooling flows in central
dominant galaxies, but we recognise that extending the wavelengh
baseline of the observations is necessary to provide stronger
constraints.  Moreover, a systematic study of the spatial extent and
morphology of the star formation (e.g., via H$\alpha$ images) and its
correlation with radio emission would provide essential information on
the suggested link between radio activity and star-formation.
Near-infrared images would add to the picture by characterising the
distribution of the underlying old stars.

Progress is also required in the detailed modelling of the behaviour of
the spectral features for potentially complex stellar population with a
mixture of ages, metal abundances and element ratios.  A missing
ingredient is a reliable empirical calibration of some spectral
features with intrinsic stellar parameters.  At present we are near to
complete the D$_{4000}$ calibration in the Lick stellar library 
(Cardiel, Gorgas \& Pedraz 1998).

\section*{Acknowledgments}

Valuable discussions with Steven Allen, Carolin Crawford, Alastair Edge, Andy
Fabian, Roderick Johnstone, Paul Nulsen, Jes\'{u}s Gallego and 
Armando Gil de Paz are gratefully acknowledged. We
thank Brian McNamara for a careful reading of the manuscript, and
Clovis Peres for kindly providing the mass deposition rates and cooling
radii employed in this paper prior to their publication. NCL would like to 
thank the Institute of Astronomy, Cambridge, for its hospitality.
The WHT is operated on the island of La Palma by the
Royal Greenwich Observatory at the Observatorio del Roque de los Muchachos of
the Instituto de Astrof\'{\i}sica de Canarias. The Calar Alto Observatory is
operated jointly by the Max-Planck-Institute f\"{u}r Astronomie, Heidelberg,
and the Spanish Comisi\'{o}n Nacional de Astronom\'{\i}a. This work was
supported in part by the Spanish `Programa Sectorial de Promoci\'{o}n del
Conocimiento' under grant \mbox{No.~PB96-610}.
AAS acknowledges generous financial support from the Royal Society.


\appendix
\section{Radial profile of the star formation}

We derive the spatial profile of the density of the mass transformed
into stars (Fig.~14) assuming that the change of slope
observed in the D$_{4000}$ gradients in CCGs with cooling flows and emission
lines are due, exclusively, to star formation (see
Section~\ref{sectiondiscussiongradients}).  In order to avoid
uncertainties in the absolute values of the total V~luminosity of the CCGs in
the emission-line regions, we only derive the relative radial
variations (normalizing both the density of the mass transformed into stars
and the spatial extent of the emission-line region).

The extrapolation of the outer D$_{4000}$ gradients into the emission-line
region allows us to estimate the expected radial behaviour of the D$_{4000}$
corresponding to the underlying old stellar population.  Assuming a
particular star formation model (in this case we have chosen a single burst
with an age $\simeq 0.1$~Gyr, with solar metallicity and Scalo IMF), it is
possible to derive the $f_V$ required to reproduce the observed D$_{4000}$
decrease (i.e. the difference between the measured and the extrapolated
D$_{4000}$ values) as a function of radius.

If we assume that galaxies exhibit spherical geometry, 
for any concentric shell Eq.~\ref{eqburst} can be written as
\begin{equation}
M_{\rm AP}(k) = f_V(k) \; L_V(k) \; (M/L)_{\rm AP},
\label{eqapenmap}
\end{equation}
where $k$ is an integer which indicates the number of the shell
(numbered from the center to outside),
$M_{\rm AP}(k)$ and $f_V(k)$ are the total mass transformed into stars and the 
fraction of V~light that comes from the accretion population in the 
\mbox{$k^{\rm th}$-shell} (respectively), 
and $(M/L)_{\rm AP}$ the mass-to-light ratio of the 
accretion population given by the adopted stellar population model.
The total V~luminosity of the \mbox{$k^{\rm th}$-shell} (in arbitrary units) is
computed from the number of counts in the V~wavelength range within
the spectrograph slit, $N_V(k)$, at a radius $r(k)$ (in arbitrary units) using
\begin{equation}
L_V(k) = N_V(k) \; r(k).
\end{equation}
Since the mass-to-light ratio of the accretion population is a constant factor
and we are only interested in relative values, we can
re-write Eq.~\ref{eqapenmap} as
\begin{equation}
M_{\rm AP}(k) = f_V(k) \; L_V(k),
\end{equation}
where both $f_V(k)$ and $L_V(k)$ can be readily evaluated for each $k$~shell.

In order to obtain a realistic three-dimensional radial profile, it is
necessary to deproject the measured two-dimensional radial variation.  The
followed deprojection algorithm is based on the computation from outside to
the center. In this sense, the first $M_{\rm AP}(k)$ computed corresponds to
the most external value ($k=N_{\rm shells}$, being $N_{\rm shells}$ the total
number of concentric shells), which is not corrected for any projection
effect. This
initial value is then employed to obtain $\rho_{\rm AP}(N_{\rm shells})$ by
dividing $M_{\rm AP}(N_{\rm shells})$ by the volume observed in projection in
the external shell (in arbitrary units). In the successive steps we move towards
the central regions (decreasing $k$), making use of the already determined
$\rho_{\rm AP}(i)$ values (with $k < i \leq N_{\rm shells}$). In addition
\begin{equation}
M_{\rm AP}(k) = \sum_{i=k}^{N_{\rm shells}} \rho_{\rm AP}(i) \; V(k,i),
\end{equation}
where $V(k,i)$ is the volume of the $i^{\rm th}$-shell in the line of sight
of the $k^{\rm th}$-shell, with \mbox{$k \leq i \leq N_{\rm shells}$} (see
Fig.~A1 and figure caption). Finally, we compute the desired
spatial profile employing
\begin{equation}
\rho_{\rm AP}(k) = 
 \frac{1}{V(k,k)} 
 \left(
    M_{\rm AP}(k) - \sum_{i=k+1}^{N_{\rm shells}} \rho_{\rm AP}(i) \; V(k,i)
  \right).
\end{equation}


\clearpage

\def\thetable{\arabic{table}}

\begin{table}
 \caption{Observational configuration.}
 \begin{tabular}{@{}lcc}
                     & Run 1              & Run 2             \\
                     &                    &                   \\
 Dates               & 9--10 Aug 1994     & 17--19 Dec 1995   \\
 Telescope           & CAHA 3.5m          & WHT 4.2m          \\
 Spectrograph        & CTS                & ISIS blue arm     \\
 Detector            & CCD TEK~12         & CCD TEK~1         \\
 Dispersion          & 3.46\ \AA/pixel    & 2.90\ \AA/pixel   \\
 Wavelength Range    & 3700--7240\ \AA    & 3740--6700\ \AA   \\
 Spectral Resolution & 8.1\ \AA\ ({\sc FWHM}) & 12.3\ \AA\ ({\sc FWHM})  \\
 Slit Width          & 2.1 arcsec         & 2.0 arcsec        \\
 Spatial Scale       & 0.9\ arcsec/pixel  & 1.1 arcsec/pixel  \\
 \end{tabular}
\end{table}

\clearpage

\begin{table*}
\begin{minipage}{114mm}
 \caption{Galaxy sample.}
 \begin{tabular}{@{}lccccccr}
 Cluster  & Galaxy   & R.A.                & Decl.                & z      
& Run & Exposure                & P.A.     \\
          &          & (1950)              & (1950)               &        
&     & (secs)                  & \hfill (\degr) \hfill \\
 A~85     & \ldots   & \cooa{00}{39}{18.7} & $-$\cood{09}{34}{41} & 0.0556 
& 2   & \makebox[8mm][r]{ 7000} & 145 \\
 A~478    & \ldots   & \cooa{04}{10}{40.7} & $+$\cood{10}{20}{20} & 0.0859 
& 2   & \makebox[8mm][r]{14000} &  55 \\
 A~496    & \ldots   & \cooa{04}{31}{18.7} & $-$\cood{13}{21}{56} & 0.0330 
& 2   & \makebox[8mm][r]{ 9000} & 175 \\
 A~644    & \ldots   & \cooa{08}{14}{59.0} & $-$\cood{07}{21}{23} & 0.0707 
& 2   & \makebox[8mm][r]{ 9000} &  16 \\
 Hydra~A  & \ldots   & \cooa{09}{15}{41.2} & $-$\cood{11}{53}{05} & 0.0544 
& 2   & \makebox[8mm][r]{ 8575} & 170 \\
 A~779    & N~2832   & \cooa{09}{16}{44.0} & $+$\cood{33}{57}{40} & 0.0229 
& 2   & \makebox[8mm][r]{ 7500} & 160 \\
 A~1377   & \ldots   & \cooa{11}{44}{41.6} & $+$\cood{56}{00}{28} & 0.0514 
& 2   & \makebox[8mm][r]{ 1800} &  60 \\
 A~2142   & \ldots   & \cooa{15}{56}{16.0} & $+$\cood{27}{22}{33} & 0.0907 
& 1   & \makebox[8mm][r]{15000} & 122 \\
 A~2255   & \ldots   & \cooa{17}{12}{09.7} & $+$\cood{64}{07}{06} & 0.0734 
& 1   & \makebox[8mm][r]{12700} &  45 \\
 A~2597   & \ldots   & \cooa{23}{22}{43.7} & $-$\cood{12}{24}{00} & 0.0826 
& 2   & \makebox[8mm][r]{ 3000} & 140 \\
 A~2626   & \ldots   & \cooa{23}{33}{59.5} & $+$\cood{20}{52}{08} & 0.0552 
& 1   & \makebox[8mm][r]{12000} &  42 \\
 \end{tabular}
\end{minipage}
\end{table*}

\clearpage

\begin{table*}
\begin{minipage}{150mm}
 \caption{Central D$_{4000}$ and Mg$_2$ indices, and associated random
errors (\mbox{$\Delta$D$_{4000}$} and \mbox{$\Delta$Mg$_2$}), for the
central galaxies of the clusters named in the first column.  The
measurements correspond to a fixed metric aperture of 4~arcsec at the
distance of the Coma cluster ($\simeq2.6\,$kpc). We have included the
sample of CGA95, with revised central measurements for A~1795 and
A~2199 (due to emission line contamination). The Mg$_2$ values have
been converted to the Lick system using the systematic offsets
derived from the observation of template stars (see text).  Mass
deposition rate, $\dot{M}$, and cooling radius, $r_{\rm cool}$, for
each cluster are taken from: [1] Peres et al. (1997, PSPC data), [2]
Edge et al. (1992), [3] MO92, [4] HBvBM, and [5] Arnaud (1988). When
available, the values quoted for these last two parameters are the
median, 10 and 90 percentile estimates.  Central mass deposition rates
(i.e.  corresponding to the same metric aperture than that employed in
the measurement of the line-strength indices) have been obtained
assuming $\dot{M} \propto r$. The $r_{\rm cool}$ value for A~1126
is the mean cooling radii for the four galaxies with similar mass
deposition rate (A~85, A~644, Hydra~A and A~2597).}
  \begin{tabular}{@{}lcccccccc}
  Cluster &
  D$_{4000}$ & $\Delta$D$_{4000}$ & Mg$_2$ & $\Delta$Mg$_2$ &
  $ \dot{M}$     & $r_{\rm cool}$ & central $\dot{M}$ & source of \\
          &
    &                    & (mag)  & \raisebox{0.ex}[1.5ex][2.5ex]{(mag)} &
  (M$_\odot$/yr) & (kpc)          & (M$_\odot$/yr)    & 
 $\dot{M}$, $r_{\rm cool}$ \\
\namelarge{A 85    } &
2.09 & 0.03 & 0.300 & 0.008 &
\mdot{ 198}{  53}{  52} &
\rcool{146}{ 41}{ 41} &
\mdot{ 1.75}{ 0.68}{ 0.67} & 1 \\
\namelarge{A 262   } &
2.34 & 0.04 & 0.330 & 0.009 &
\mdot{  27}{   4}{   3} &
\rcool{104}{ 11}{ 10} &
\mdot{ 0.33}{ 0.06}{ 0.05} & 1 \\
\namelarge{2A 0335+096 } &
2.08 & 0.04 & 0.293 & 0.009 &
\mdot{ 325}{  32}{  43} &
\rcool{215}{ 29}{ 29} &
\mdot{ 1.95}{ 0.32}{ 0.37} & 1 \\
\namelarge{A 478   } &
1.84 & 0.03 & 0.277 & 0.013 &
\mdot{ 616}{  63}{  76} &
\rcool{204}{ 27}{ 38} &
\mdot{ 3.89}{ 0.65}{ 0.87} & 1 \\
\namelarge{A 496   } &
2.25 & 0.04 & 0.318 & 0.003 &
\mdot{  95}{  13}{  12} &
\rcool{110}{ 12}{ 15} &
\mdot{ 1.11}{ 0.19}{ 0.20} & 1 \\
\namelarge{PKS 0745-191} &
1.31 & 0.04 & 0.130 & 0.045 &
\mdot{1038}{ 116}{  68} &
\rcool{214}{ 49}{ 25} &
\mdot{ 6.26}{ 1.59}{ 0.84} & 1 \\
\namelarge{A 644   } &
2.23 & 0.03 & 0.329 & 0.007 &
\mdot{ 189}{ 106}{  35} &
\rcool{141}{ 62}{ 18} &
\mdot{ 1.73}{ 1.23}{ 0.39} & 1 \\
\namelarge{Hydra A     } &
1.44 & 0.02 & 0.246 & 0.004 &
\mdot{ 264}{  81}{  60} &
\rcool{162}{ 56}{ 68} &
\mdot{ 2.10}{ 0.97}{ 1.00} & 1 \\
\namelarge{A 779   } &
2.28 & 0.04 & 0.347 & 0.002 &
\mdotvoid{   0}{    }{    } &
\ldotss &
\ldotss & 3 \\
\namelarge{A 1126  } &
1.82 & 0.03 & 0.294 & 0.009 &
\mdotvoid{ 222}{    }{    } &
\rcool{150}{ 50}{ 50} &
\mdot{ 1.91}{ 0.64}{ 0.64} & 4 \\
\namelarge{A 1377  } &
2.26 & 0.04 & 0.321 & 0.009 &
\mdotvoid{   0}{    }{    } &
\ldotss &
\ldotss & 5 \\
\namelarge{A 1795  } &
1.63 & 0.03 & 0.249 & 0.010 &
\mdot{ 381}{  41}{  23} &
\rcool{177}{ 19}{  6} &
\mdot{ 2.78}{ 0.42}{ 0.19} & 1 \\
\namelarge{A 2124  } &
2.16 & 0.04 & 0.293 & 0.008 &
\mdotvoid{   0}{    }{    } &
\ldotss &
\ldotss & 5 \\
\namelarge{A 2142  } &
2.36 & 0.05 & 0.334 & 0.006 &
\mdot{ 350}{  66}{ 133} &
\rcool{150}{ 18}{ 49} &
\mdot{ 3.01}{ 0.67}{ 1.51} & 1 \\
\namelarge{A 2199  } &
2.10 & 0.05 & 0.322 & 0.007 &
\mdot{ 154}{  18}{   8} &
\rcool{143}{ 17}{  6} &
\mdot{ 1.39}{ 0.23}{ 0.09} & 1 \\
\namelarge{A 2255  } &
2.42 & 0.03 & 0.306 & 0.017 &
\mdotvoid{   0}{    }{    } &
\ldotss &
\ldotss & 2 \\
\namelarge{A 2319  } &
2.02 & 0.04 & 0.311 & 0.011 &
\mdot{  20}{  61}{  20} &
\rcool{ 53}{ 59}{ 53} &
\mdot{ 0.47}{ 1.58}{ 0.67} & 1 \\
\namelarge{A 2597  } &
1.64 & 0.02 & 0.260 & 0.016 &
\mdot{ 271}{  41}{  41} &
\rcool{152}{ 67}{ 58} &
\mdot{ 2.30}{ 1.07}{ 0.94} & 1 \\
\namelarge{A 2626  } &
2.16 & 0.03 & 0.291 & 0.013 &
\mdot{  36}{  16}{  36} &
\rcool{119}{ 60}{ 60} &
\mdot{ 0.38}{ 0.26}{ 0.43} & 5 \\
\namelarge{A 2634  } &
2.21 & 0.02 & 0.300 & 0.007 &
\mdotvoid{   0}{    }{    } &
\ldotss &
\ldotss & 5 \\
  \end{tabular}
\end{minipage}
\end{table*}

\clearpage

\begin{table*}
\begin{minipage}{165mm}
\caption{D$_{4000}$ and Mg$_2$ gradients
(${\rm d D}_{4000}/{\rm d}\log r$, ${\rm d Mg}_2/{\rm d}\log r$),
corresponding to the error-weighted
least-squares fits shown in Figs.~3 and~4.
Total gradients refer to fits computed by employing all the data excluding
measurements within 1.5~arcsec and secondary nuclei. Inner and outer gradients
correspond to fits inside and outside the emission line region, when present.
Formal errors of the fits are given next to each gradient.}
\begin{tabular}{@{}l@{$\;\;\;$}c@{$\;\;\;$}c@{$\;$}c@{$\;\;\;$}c@{$\;$}%
c@{$\;\;\;$}c@{$\;$}c@{$\;\;\;$}c@{$\;$}c@{$\;\;\;$}c@{$\;$}c@{$\;\;\;$}%
c@{$\;$}c}
        & emission & 
\multicolumn{2}{c}{total} & \multicolumn{2}{c}{inner} & 
\multicolumn{2}{c}{outer} &
\multicolumn{2}{c}{total} & \multicolumn{2}{c}{inner} &
\multicolumn{2}{c}{outer} \\
Cluster &  lines   & 
G$_{{\rm D}_{4000}}$ & $\Delta$G$_{{\rm D}_{4000}}$ &
G$_{{\rm D}_{4000}}$ & $\Delta$G$_{{\rm D}_{4000}}$ &
G$_{{\rm D}_{4000}}$ & $\Delta$G$_{{\rm D}_{4000}}$ &
G$_{{\rm Mg}_{2}}$ & $\Delta$G$_{{\rm Mg}_{2}}$ &
G$_{{\rm Mg}_{2}}$ & $\Delta$G$_{{\rm Mg}_{2}}$ &
G$_{{\rm Mg}_{2}}$ & $\Delta$G$_{{\rm Mg}_{2}}$ \\
 & & & & & & & & & & & & & \\
A85     & yes & 
$-$0.134 & 0.031 & $+$0.044 & 0.025 & $-$0.312 & 0.025 &
$-$0.005 & 0.007 & $+$0.007 & 0.014 & $-$0.030 & 0.014 \\
A262    & yes & 
$-$0.142 & 0.037 & $-$0.166 & 0.039 & $+$0.111 & 0.298 &
$-$0.021 & 0.011 & $-$0.017 & 0.013 & $-$0.085 & 0.066 \\
A478    & yes & 
$+$0.239 & 0.109 & $+$0.389 & 0.137 & $-$0.313 & 0.168 &
 \ldotss & \ldotss  & $+$0.015 & 0.025 & \ldotss  & \ldotss  \\
A496    & yes & 
$-$0.139 & 0.023 & $-$0.112 & 0.023 & $-$0.205 & 0.043 &
$-$0.036 & 0.008 & $-$0.022 & 0.010 & $-$0.085 & 0.016 \\
A644    & no  & 
$-$0.211 & 0.063 & \ldotss  & \ldotss  & \ldotss  & \ldotss  &
$-$0.043 & 0.013 & \ldotss  & \ldotss  & \ldotss  & \ldotss  \\
Hydra A & yes & 
$+$0.254 & 0.094 & $+$0.536 & 0.036 & $-$1.018 & 0.063 &
$+$0.046 & 0.010 & $+$0.048 & 0.010 & $-$0.030 & 0.097 \\
A779    & no  & 
$-$0.246 & 0.010 & \ldotss  & \ldotss  & \ldotss  & \ldotss  &
$-$0.065 & 0.005 & \ldotss  & \ldotss  & \ldotss  & \ldotss  \\
A1126   & yes & 
$-$0.114 & 0.067 & $-$0.206 & 0.074 & $+$0.099 & 0.088 &
$-$0.045 & 0.011 & $-$0.059 & 0.014 & $+$0.006 & 0.046 \\
A1377   & no  & 
$-$0.219 & 0.097 & \ldotss  & \ldotss  & \ldotss  & \ldotss  &
$+$0.001 & 0.020 & \ldotss  & \ldotss  & \ldotss  & \ldotss  \\
A1795   & yes & 
$-$0.162 & 0.046 & $-$0.002 & 0.043 & $-$0.501 & 0.044 &
$-$0.079 & 0.012 & $-$0.052 & 0.009 & $-$0.192 & 0.032 \\
A2124   & no  & 
$-$0.425 & 0.052 & \ldotss  & \ldotss  & \ldotss  & \ldotss  &
$-$0.107 & 0.014 & \ldotss  & \ldotss  & \ldotss  & \ldotss  \\
A2142   & no  & 
$-$0.312 & 0.034 & \ldotss  & \ldotss  & \ldotss  & \ldotss  &
$-$0.061 & 0.009 & \ldotss  & \ldotss  & \ldotss  & \ldotss  \\
A2199   & yes & 
$-$0.397 & 0.036 & $-$0.129 & 0.128 & $-$0.448 & 0.023 &
$-$0.083 & 0.013 & $-$0.015 & 0.006 & $-$0.126 & 0.008 \\
A2255   & no  & 
$-$0.286 & 0.073 & \ldotss  & \ldotss  & \ldotss  & \ldotss  &
$-$0.060 & 0.013 & \ldotss  & \ldotss  & \ldotss  & \ldotss  \\
A2319   & no  & 
$-$0.418 & 0.062 & \ldotss  & \ldotss  & \ldotss  & \ldotss  &
$-$0.052 & 0.021 & \ldotss  & \ldotss  & \ldotss  & \ldotss  \\
A2597   & yes & 
$+$0.112 & 0.068 & $+$0.189 & 0.053 & $-$0.523 & 0.123 & 
 \ldotss & \ldotss  & $+$0.038 & 0.024 & \ldotss  & \ldotss  \\
A2626   & yes & 
$-$0.376 & 0.059 & $-$0.236 & 0.027 & $-$0.487 & 0.033 &
$-$0.048 & 0.007 & $-$0.019 & 0.008 & $-$0.076 & 0.012 \\
A2634   & yes & 
$-$0.219 & 0.130 & \ldotss  & \ldotss  & \ldotss  & \ldotss  &
$-$0.073 & 0.032 & \ldotss  & \ldotss  & \ldotss  & \ldotss  \\
\end{tabular}
\end{minipage}
\end{table*}

\clearpage

\begin{table*}
\begin{minipage}{132mm}
 \caption{Central D$_{4000}$ indices derived from the nuclear \mbox{(U$-$b)} 
colours published by MO92, using the linear relation fitted in 
Fig.~5b. Errors ($\Delta {\rm D}_{4000}$) have been 
derived from the (U$-$b) errors reported by MO92.
Sources for mass deposition rate and 
cooling radius are those quoted in Table~3.}
 \begin{tabular}{@{}lccccccc}
Cluster & Galaxy & D$_{4000}$ & $\Delta {\rm D}_{4000}$ &
$\dot{M}$      & $r_{\rm cool}$ & central $\dot{M}$ & 
source of \\
  &  &  &  & (M$_\odot$/yr) & 
\raisebox{0.ex}[1.5ex][2.5ex]{(kpc)}   & (M$_\odot$/yr)    &
$\dot{M}$, $r_{\rm cool}$ \\
\namelarge{A~119}  & \ldots &   2.16     &           0.03          &
\mdot{   0}{  2}{  0} & 
\rcool{  0}{ 62}{  0} &
\mdot{ 0.00}{ 0.00}{ 0.00} & 1 \\
\namelarge{A~426}  & N~1275 &   1.27     &           0.03          &
\mdot{ 556}{ 33}{ 24} & 
\rcool{185}{ 11}{ 11} &
\mdot{ 3.88}{ 0.23}{ 0.17} & 1 \\
\namelarge{MKW~1}  & N~3090 &   2.26     &           0.03          &
\mdotvoid{   0}{   }{   } & 
\ldotss &
\ldotss & 3 \\
\namelarge{Virgo}  &  M~87  &   1.97     &           0.03          &
\mdot{  39}{  2}{  9} & 
\rcool{102}{  6}{  4} &
\mdot{ 0.49}{ 0.03}{ 0.11} & 1 \\
\namelarge{MKW~5}  & N~5400 &   2.21     &           0.03          &
\mdotvoid{   0}{   }{   } & 
\ldotss &
\ldotss & 3 \\
\namelarge{A~1991} & N~5778 &   2.11     &           0.03          &
\mdot{  71}{ 38}{ 23} & 
\rcool{142}{ 48}{ 48} &
\mdot{ 0.65}{ 0.34}{ 0.21} & 5 \\
\namelarge{A~2029} & IC~1101&   2.18     &           0.03          &
\mdot{ 556}{ 44}{ 73} & 
\rcool{186}{ 19}{ 39} &
\mdot{ 3.86}{ 0.31}{ 0.51} & 1 \\
\namelarge{A~2052} & \ldots &   2.00     &           0.03          &
\mdot{ 125}{ 26}{  6} & 
\rcool{147}{ 53}{  3} &
\mdot{ 1.09}{ 0.22}{ 0.05} & 1 \\
 \end{tabular}
\end{minipage}
\end{table*}


\clearpage

{\bf Figure 1:} 
Comparison of our Mg$_2$ measurements with those measured by the Lick group
in a sample of~39~(run~1) and~13~(run~2) stars from the Lick library. The
error bars in the lower right corner indicate the typical random error. There
is a systematic offset of 0.021~mag (dashed line) with a rms scatter of 0.010
mag. The offsets found analysing each run independently exhibit an excellent
agreement: \hbox{$0.020$, rms~$=0.008$} and \hbox{$0.021$, rms~$=0.021$} mag
for run~1 and~2, respectively.

{\bf Figure 2:}
Nuclear spectrum of the central galaxy of the cluster Hydra~A with typical
emission lines. The arrows indicate the location of the Mg$_2$ bandpasses. A
simultaneous fit of a scaled template spectrum, a 2$^{\rm nd}$~order
polynomial and a Gaussian reproduces the local galaxy spectrum around the
central bandpass. The final fit (\mbox{polynomial$+$template}) is shown as
the thin line overplotted on the galaxy spectrum. The rms scatter of the
residuals (shown in the inner small box) is~0.017, absolutely consistent with
the mean random error in this spectral range (0.016). The pixels affected by
the [N\,{\sc i}]~$\lambda$5199 emission are replaced by this combined fit.
The [O\,{\sc iii}]~$\lambda$4959 emission in the blue bandpass is analogously
removed. Common emission lines interpolated in the D$_{4000}$ bandpasses are
[Ne\,{\sc iii}]~$\lambda$3869, [S\,{\sc ii}]~$\lambda\lambda$\mbox{4069,
4076}, and H$\delta$. This procedure is repeated in each spectrum with
emission lines.

{\bf Figure 3:}
D$_{4000}$ gradients. See explanation in section~\ref{sectiongradients}.

{\bf Figure 4:}
Mg$_2$ gradients. See explanation in section~\ref{sectiongradients}.

{\bf Figure 5:}
Panel (a): comparison of central D$_{4000}$ measurements from JFN87 ---open
circles--- and MO89 ---filled circles--- with our central D$_{4000}$ indices
for four and three galaxies in common, respectively. Panel (b): correlation
between the nuclear \mbox{(U$-$b)} colours from MO92 with the central
D$_{4000}$ values.

{\bf Figure 6:}
Comparison of Mg$_2$ gradients for four galaxies in common with FFI95. Open
symbols (squares: major axis, triangle: minor axis) are from FFI95, whereas
filled circles correspond to measurements from CGA95 and this work (major
axis, averaged at both sides of the galaxies).  The minor axis data from
FFI95 are rescaled to reproduce the same spatial scale than the major axis
data.  The upper x-scale in the plots indicates the spatial scale in
arcsecs.  The atomic Mgb measurements of FFI95 were transformed into
molecular Mg$_2$ values and then converted to the Lick system by applying an
offset of 0.014 mag.  This offset has been derived from a comparison of the
FFI95 data with a compilation of gradients by Gonz\'{a}lez \& Gorgas (1997).
Our Mg$_2$ indices were also transformed to the Lick system using the
systematic offsets determined through the observation of template stars from
the Lick library (see text). The short horizontal full lines at $r \sim
1$~arcsec in panels~(a), (b) and~(d) correspond to the central Mg$_2$
measurement given by Trager (1997).  Similarly, the short horizontal dashed
lines in panels~(b) and~(d) are the central Mg$_2$ values measured by Lucey
et al.  (1997). The dotted arrows in the inner regions of the central
galaxies of A~2199 and A~496 indicate the location of our Mg$_2$ indices if
we had measured Mgb in our spectra without removing the contamination by the
emission line [N\,{\sc i}]~$\lambda$5199, and then transformed these Mgb
values into Mg$_2$ (in the same way as we did with the FFI95 data). The
effect of the emission lines in A~2634 is completely negligible.

{\bf Figure 7:}
Central D$_{4000}$ and Mg$_2$ measurements versus X-ray derived mass
deposition rates (the sources for $\dot{M}$ are given in
Table~3). The central indices were obtained employing
a fixed metric aperture size of 4~arcsec projected at the distance of the
Coma cluster. Central mass deposition rates have been computed from the total
mass deposition rates assuming $\dot{M}\propto r$.  Error bars in mass
deposition rate indicate the 10 and 90 percentile estimates. Filled and open
symbols correspond to CCGs with and without emission lines in the central
regions, respectively.  Circles are galaxies from CGA95 and this paper,
whereas triangles indicate D$_{4000}$ values derived from nuclear
\mbox{(U$-$b)} colours published by MO92, as explained in the text.

{\bf Figure 8:}
Central D$_{4000}$ versus Mg$_2$ measurements. See explanation in 
section~\ref{sectiondiscussionindex}.

{\bf Figure 9:}
Central spectrum of three CCGs of our sample with emission lines (A~1795,
A~2597 and Hydra~A, from top to bottom, --thick line--), compared with a
scaled template spectra ($+$low-order polynomial) obtained from the central
regions of CCGs without emission lines (A~2124, A~644 and A~644, from top to
bottom, --thin line--). A marginal detection of He\,{\sc ii}~$\lambda$4686 is
present in A~1795 and in A~2597, whereas this is uncertain in the case of
Hydra~A.

{\bf Figure 10:}
Modelling of different star formation episodes overimposed on the spectrum of
an old stellar population. We have employed the GISSEL96 predictions (Bruzual
\& Charlot 1997), with solar metallicity and Scalo (1986) IMF. Circles are
relative UV fluxes from Crawford \& Fabian (1993) ---filled circles: A~1795,
open circles: A~2597---, whereas the filled square is an IR flux estimation
from the averaged nuclear (U-I) colours of A~1795 and A~2597 reported by
MO92. Panel~(a) shows the spectral energy distribution (SED) corresponding to
an old (15~Gyr) stellar population (thick line), a continuous star formation
during the last 10~Gyr (dotted line), and the final product of both stellar
components (thin line) assuming that the fraction of V~light that comes from
the continuous star formation is $f_V=0.3$. The prediction of this continuous
star formation (thin line) is compared, in the following panels, with the
final SEDs obtained by superimposing a single burst of star formation on the
spectrum of the old stellar population. In particular, we have represented
the resulting spectrum (thick line~=~15~Gyr~stellar population~+~single
burst, $f_V=0.25$) when 0.01, 0.1 and 0.2~Gyr have elapsed after the time the
burst took place (panels (b), (c) and~(d) respectively). See discussion in
the text.

{\bf Figure 11:}
Comparison of the break radius obtained through the simultaneous fit of two
straight lines forced to join at a common radius, with the radius of the
emission-line region. Break radii and their error bars correspond to the mean
and standard deviation derived in Monte Carlo simulations.  When available,
estimates of the break radius from D$_{4000}$ and Mg$_2$ line-strength
gradients have been averaged.  Emission-line radii where determined in the
spectroscopic images as the radii where the emission lines not longer
detectable (see section~4.2).

{\bf Figure 12:}
Line-strength gradients in D$_{4000}$ and Mg$_2$ for the galaxies tabulated
in Table~4 as a function of the mass deposition rate
---panels~(a) and~(b). Open symbols are total gradients in galaxies without
emission lines. Filled symbols correspond to inner gradients (those measured
in the emission line region). Error bars in mass deposition rate indicate the
10 and 90 percentile estimates.  In panels~(c) and~(d) we compare the mean
gradients as a function of the galaxy type: normal ellipticals (Es), central
dominant galaxies in clusters without cooling flow (CCGs no CF), and cooling
flow galaxies without (CFGs no EL) and with emission lines (CFGs EL). In the
latter type we also distinguish between inner and outer gradients, which
correspond to the emission line region and outside this region,
respectively.  The error bars in the last two panels are the rms scatter
around the mean values.  The mean D$_{4000}$ gradient for Es
($-0.315$,~rms~$=0.090$) has been computed from the data presented by Munn
(1992) and Davidge \& Clark (1994) (see CGA95 for further details). In
addition, the Mg$_2$ gradient for Es ($-0.055$,~rms~$=0.025$) is the one
derived by Gonz\'{a}lez \& Gorgas (1997) using published and unpublished data
for 109~galaxies with reliable Mg$_2$ profiles.

{\bf Figure 13:}
Total mass transformed into stars in the emission-line region of the central
galaxies with reliable D$_{4000}$ gradients in the outer parts.  The plotted
values have been derived using Eq.~\ref{eqburst}, and assuming that a single
burst of star formation (after 0.1~Gyr) has produced the observed D$_{4000}$
variation corresponding to the difference between the measured inner gradient
and the extrapolated outer gradient.

{\bf Figure 14:}
Spatial profile of the mass transformed into stars obtained through the
deprojection of the observed radial variations in D$_{4000}$ (see
Appendix~A). The dashed line is the expected density of the mass deposition
profile is one assumes that $\dot{M}(<r) \propto r$. The numbers in
parenthesis are the profile slopes obtained by least-squares linear fits.

{\bf Figure A1:}
We have obtained the spatial profile of the density of mass transformed into
stars $\rho_{\rm AP}(k)$ in each $k$-shell by subtracting the contribution of
the previously determined $\rho_{\rm AP}(i)$ ($k < i \leq N_{\rm shells}$).
For this purpose it is necessary to evaluate the volume of each $i$-shell in
the line of sight of the considered $k$-shell, $V(k,i)$.  If we define
$v(r_{\rm c},r_{\rm s})$ as the volume of the intersection between a cylinder
of radius $r_{\rm c}$ and a sphere of radius $r_{\rm s}$, it is easy to show
that
\mbox{$%
V(k,i)=v(r_{\rm b},r_{\rm 2})-
        v(r_{\rm a},r_{\rm 2})-
        v(r_{\rm b},r_{\rm 1})+
        v(r_{\rm a},r_{\rm 1})
$}, where 
\mbox{$%
v(r_{\rm c},r_{\rm s}) = 
  2 \pi r_{\rm c}^2 h +
  \frac{2}{3} \pi (r_{\rm s}-h)^2 (2 r_{\rm s} + h)
$}, being $h^2=r_{\rm s}^2-r_{\rm c}^2$.

\label{lastpage}

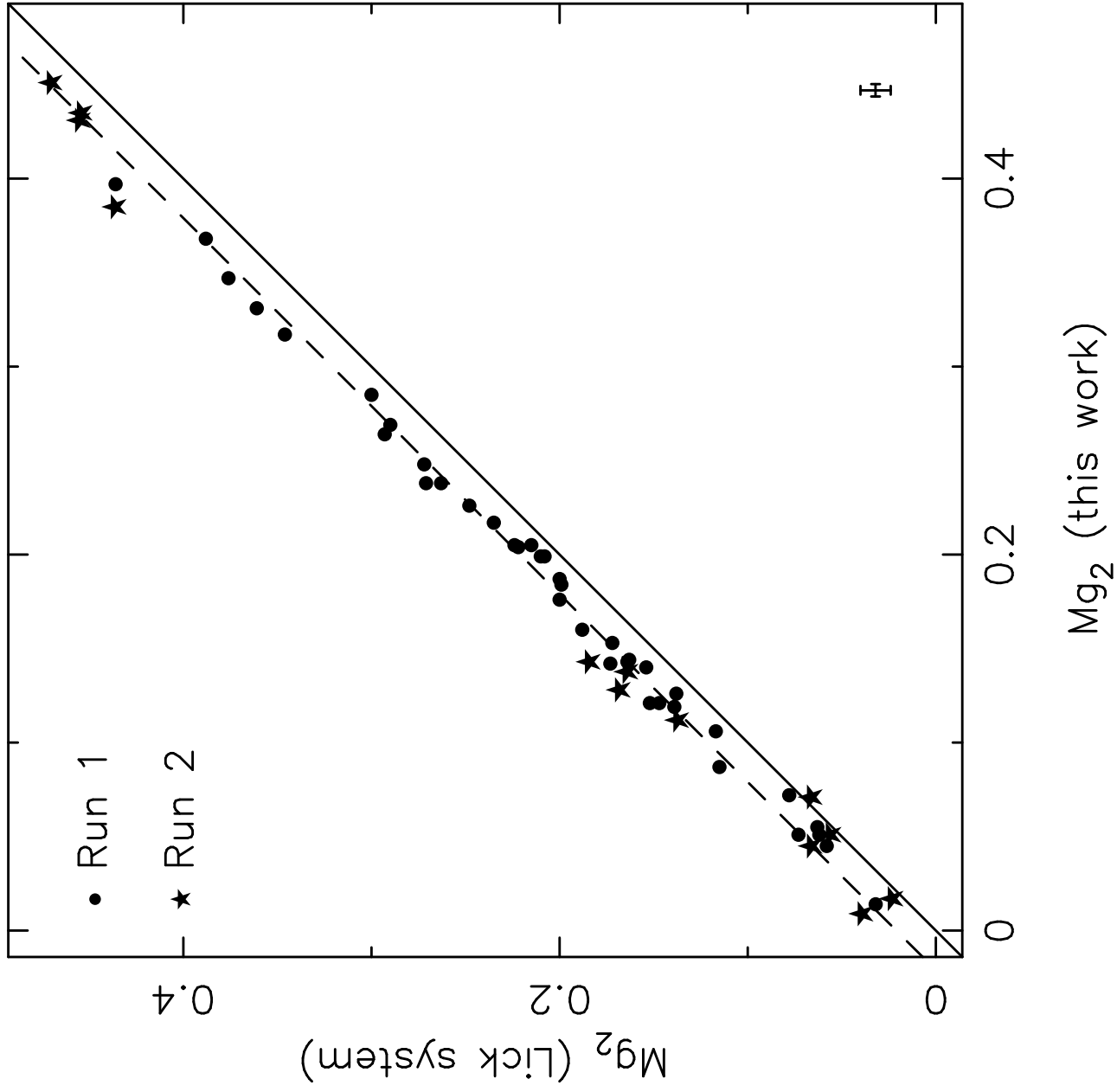
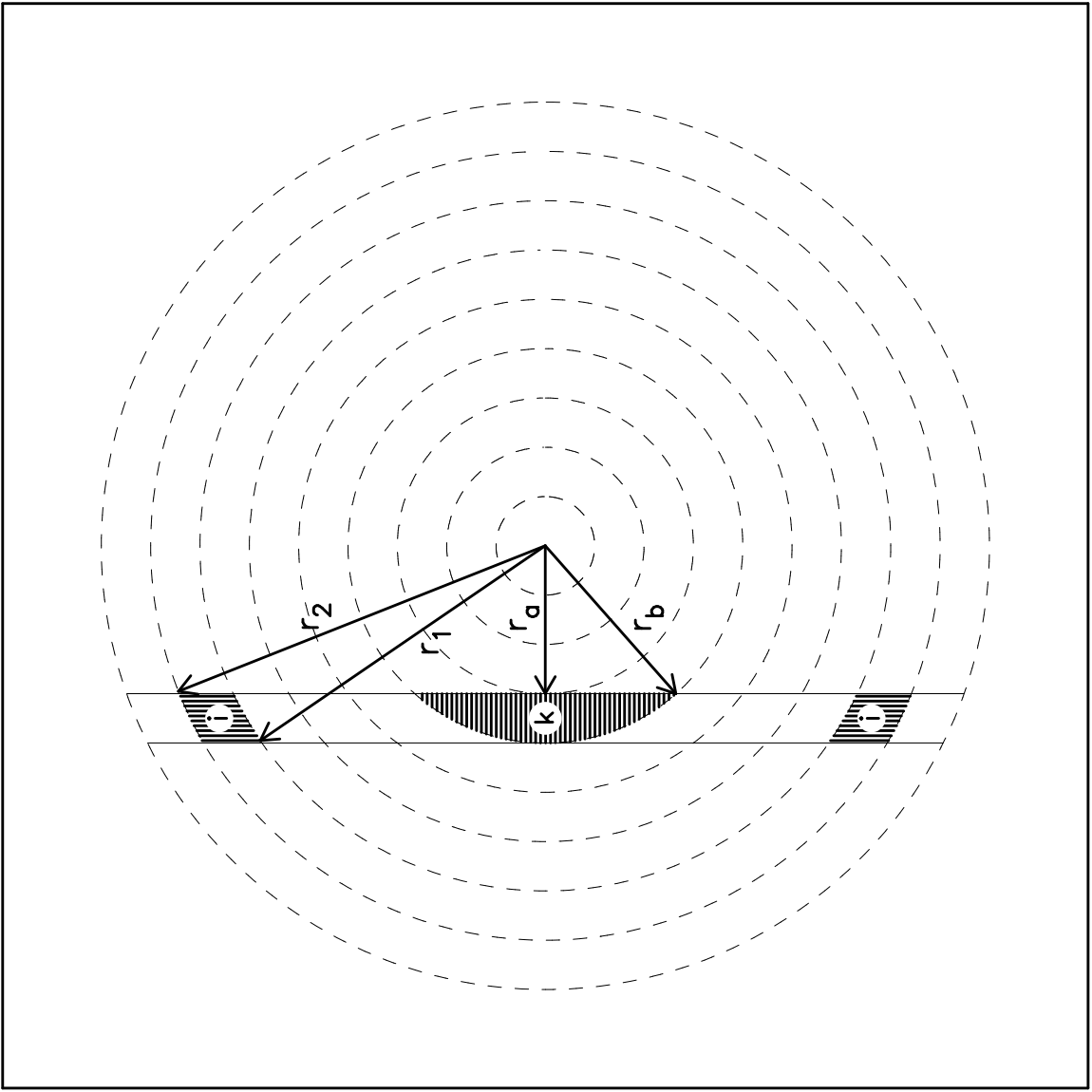
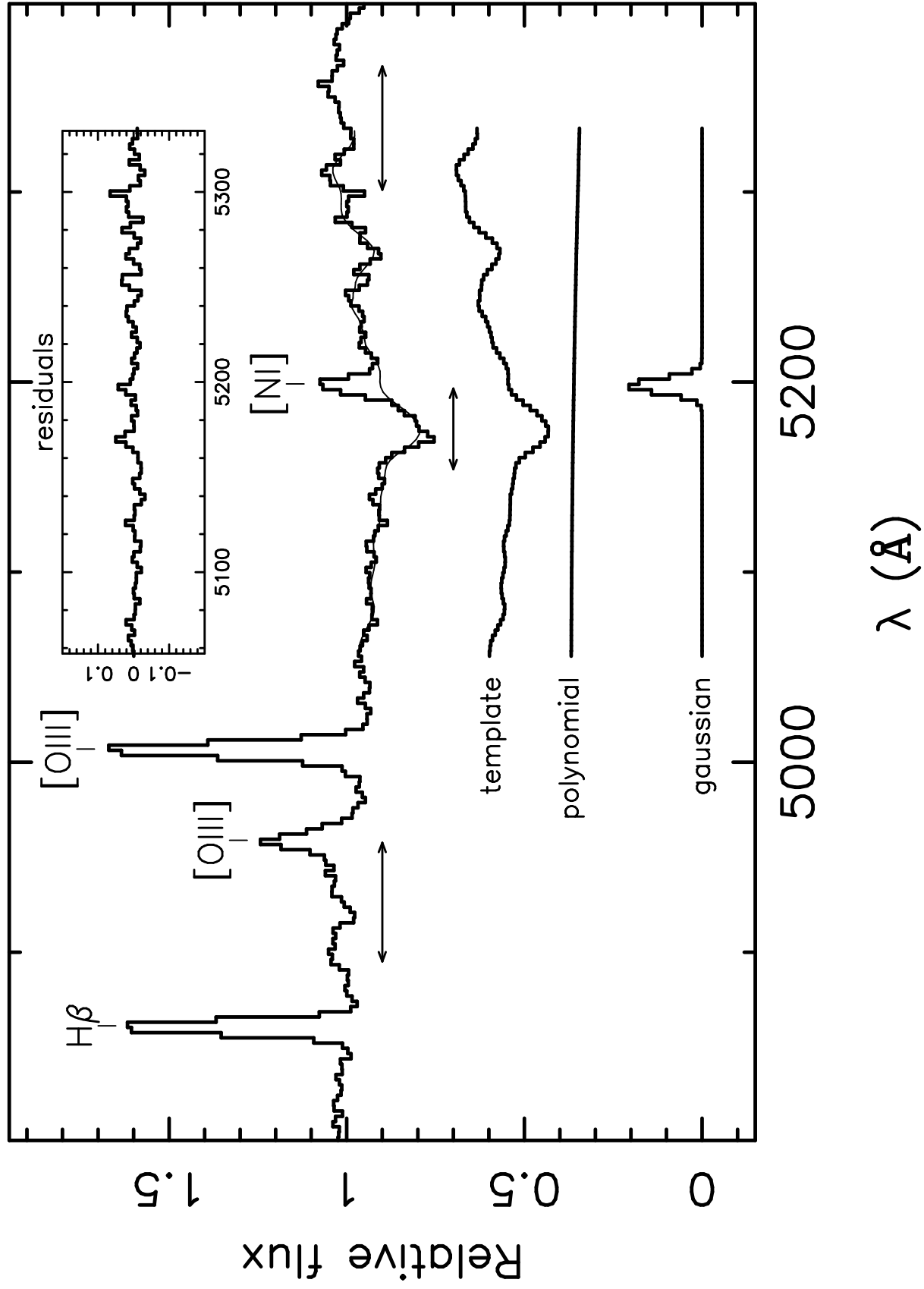
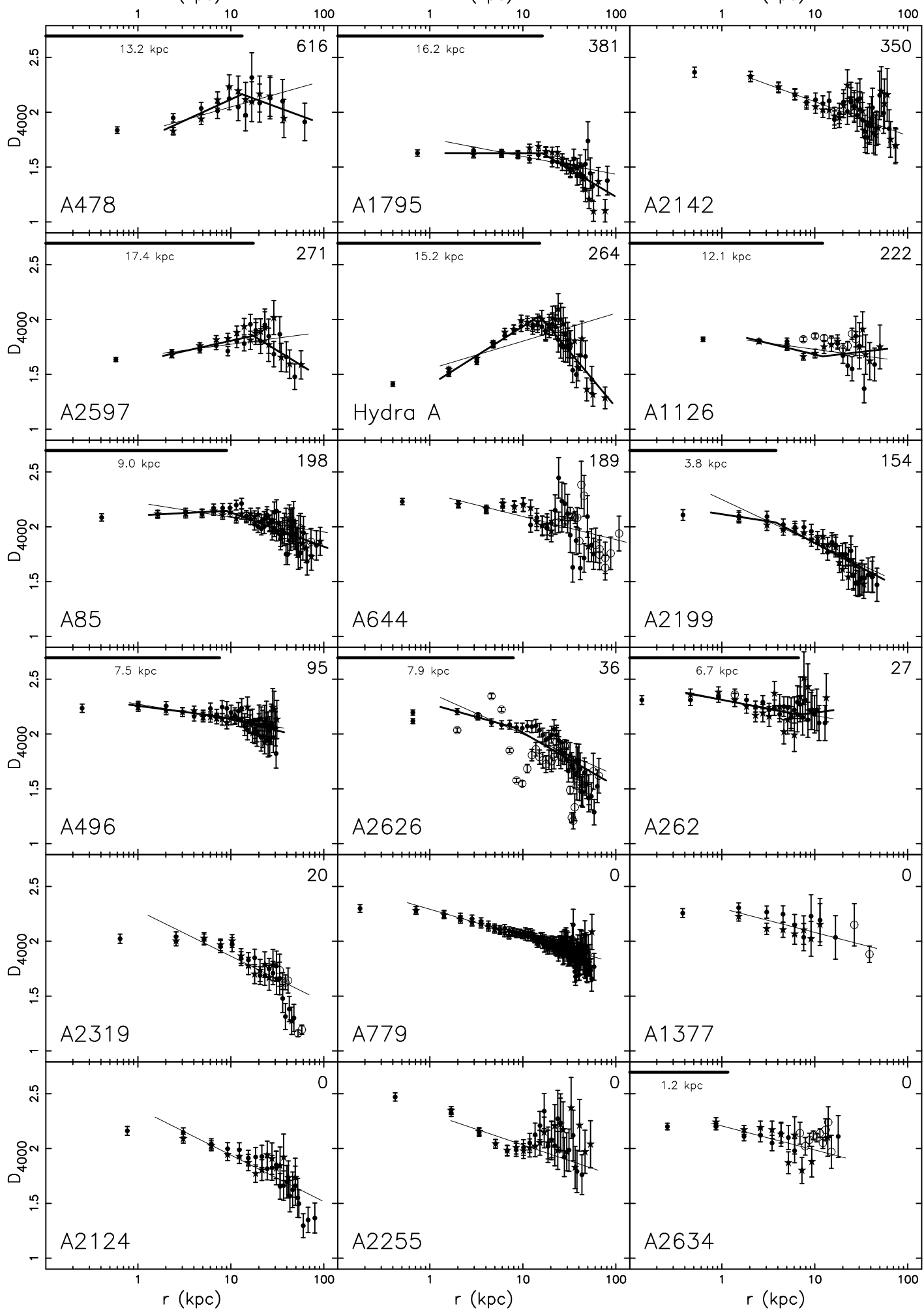
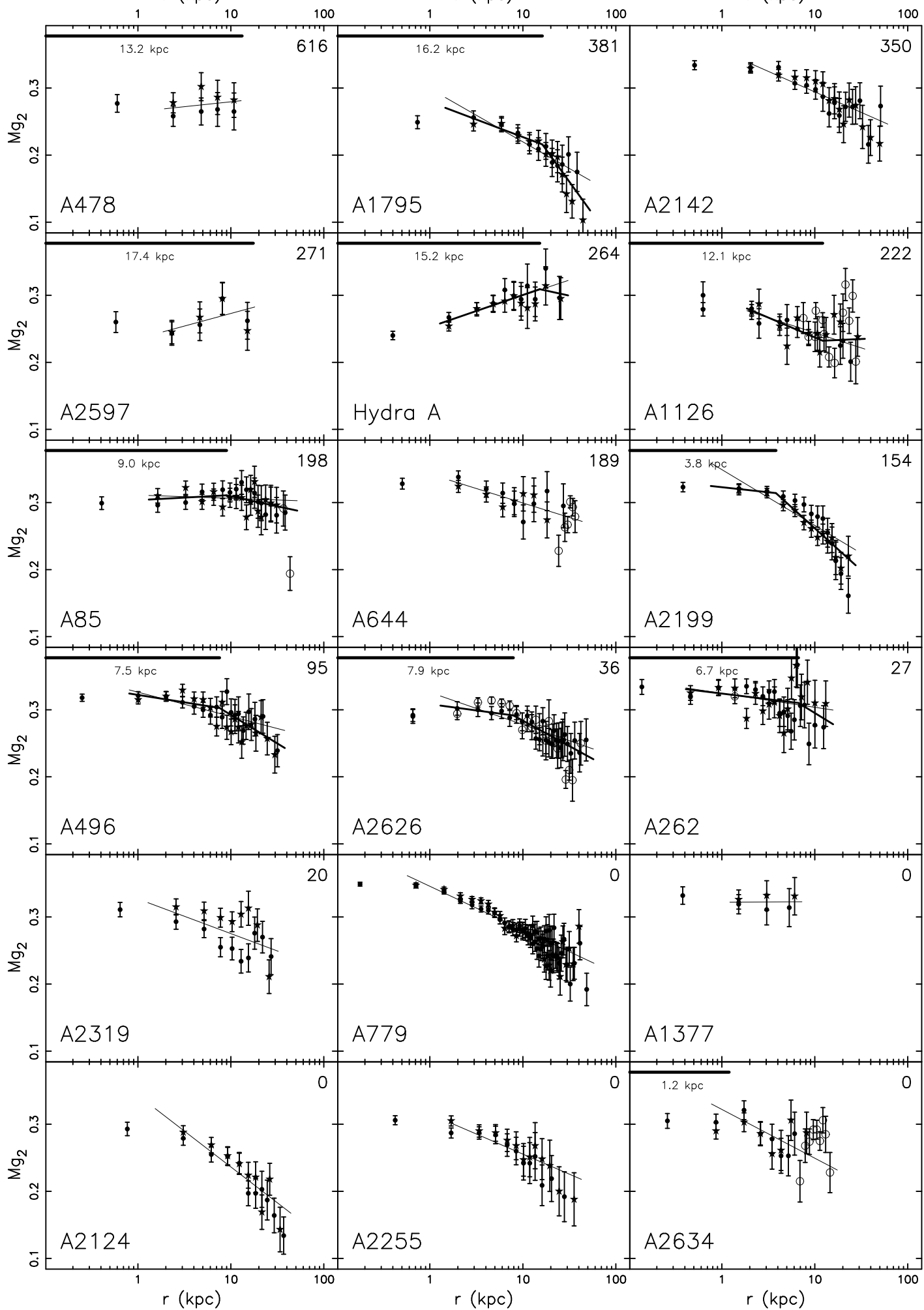
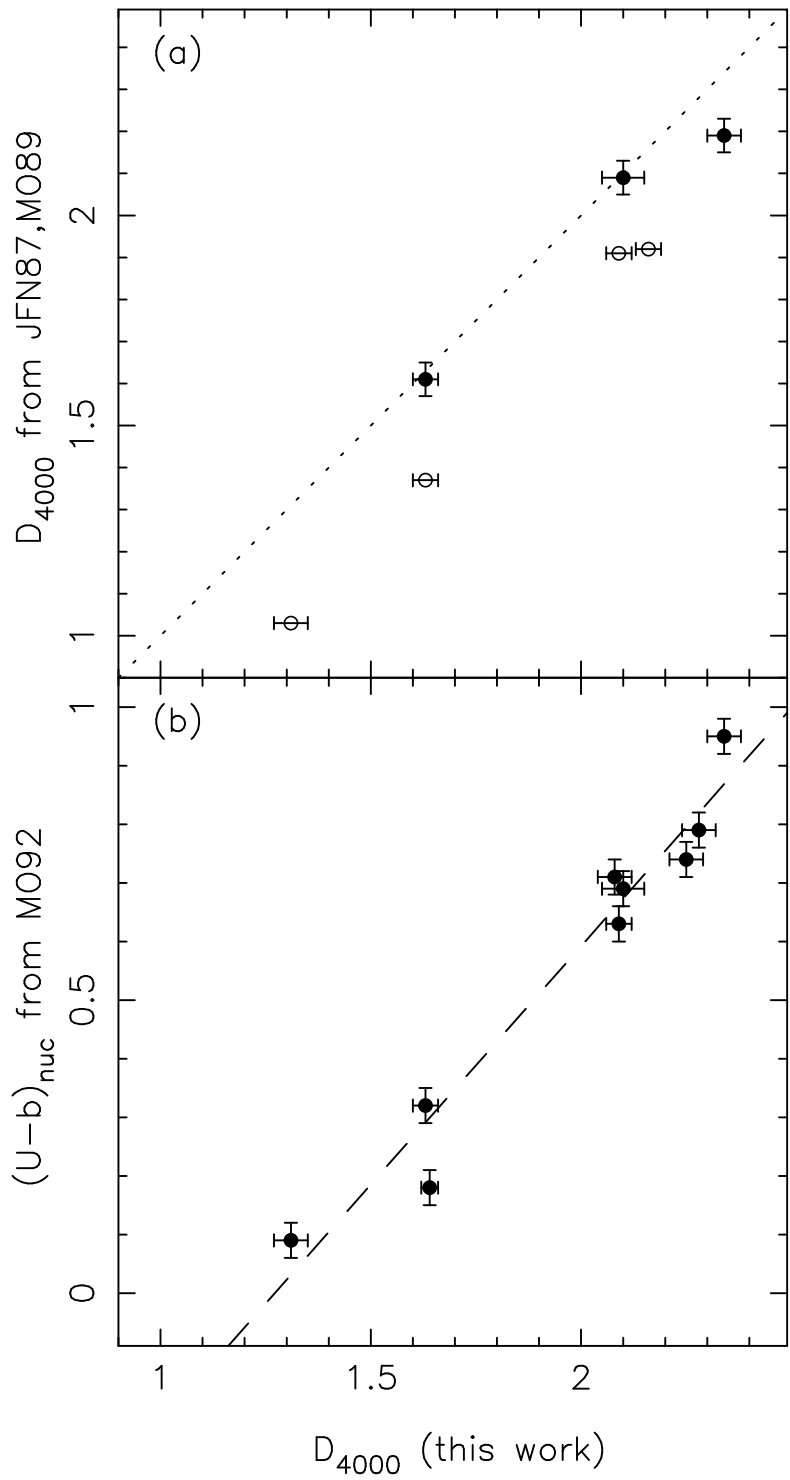
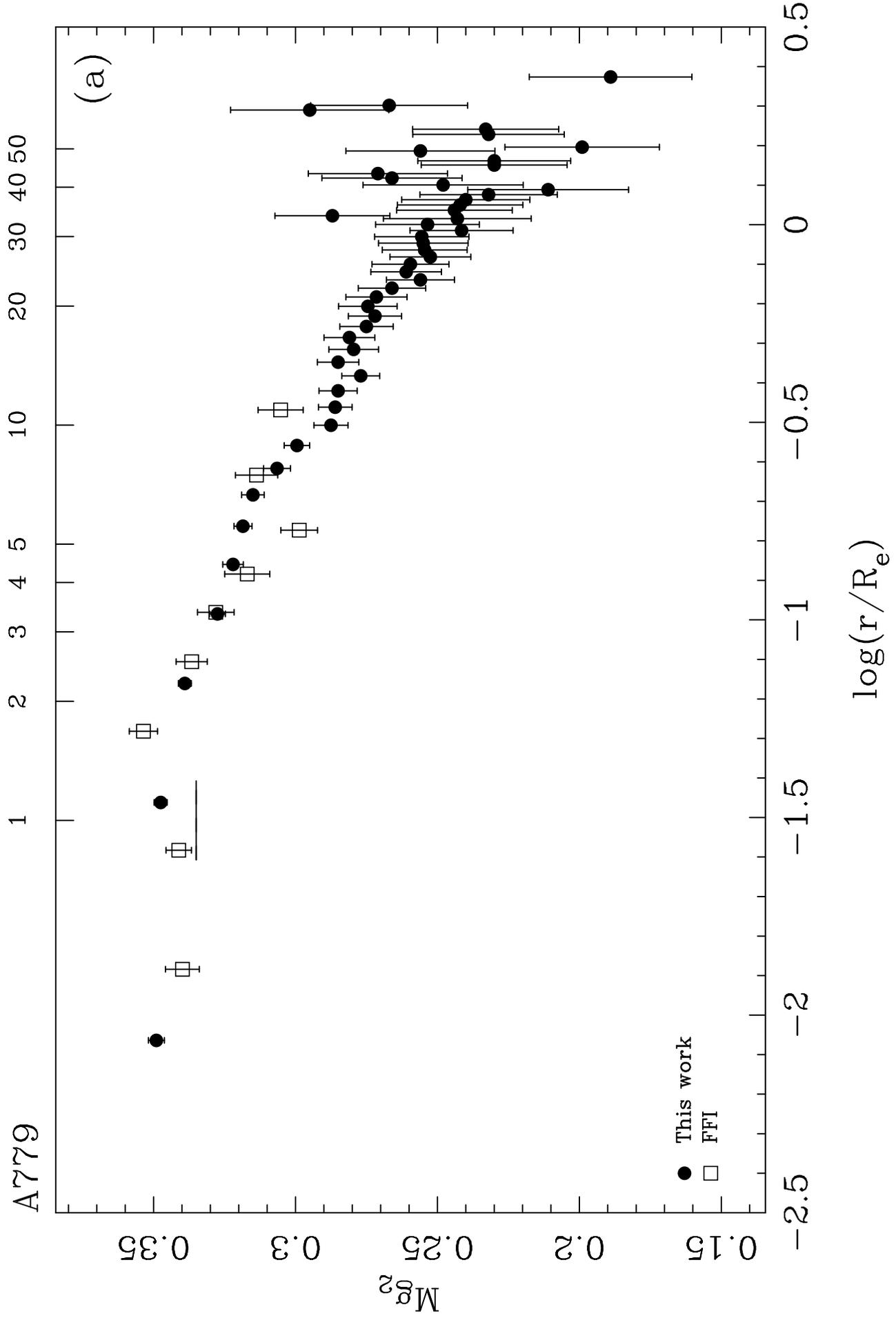
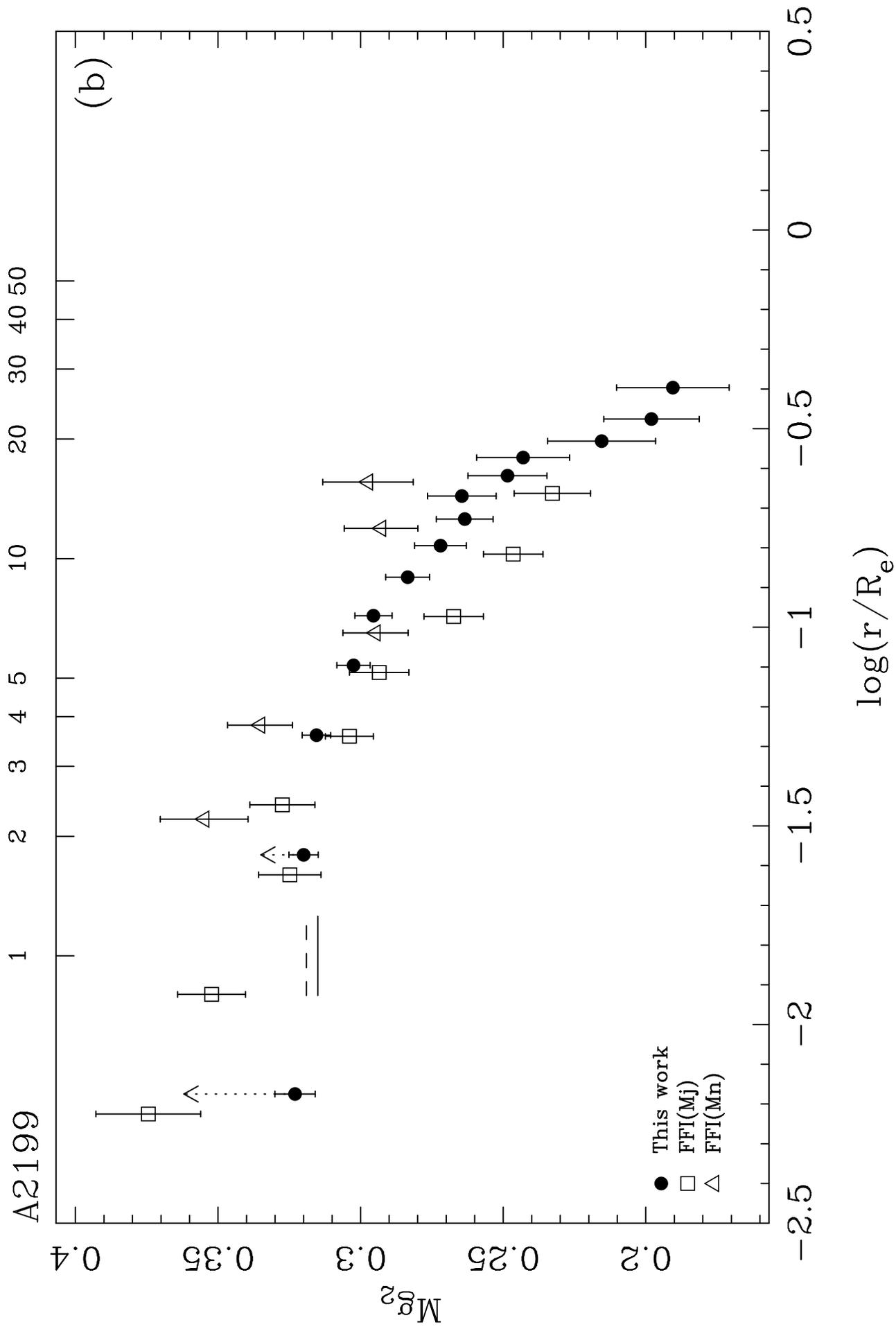
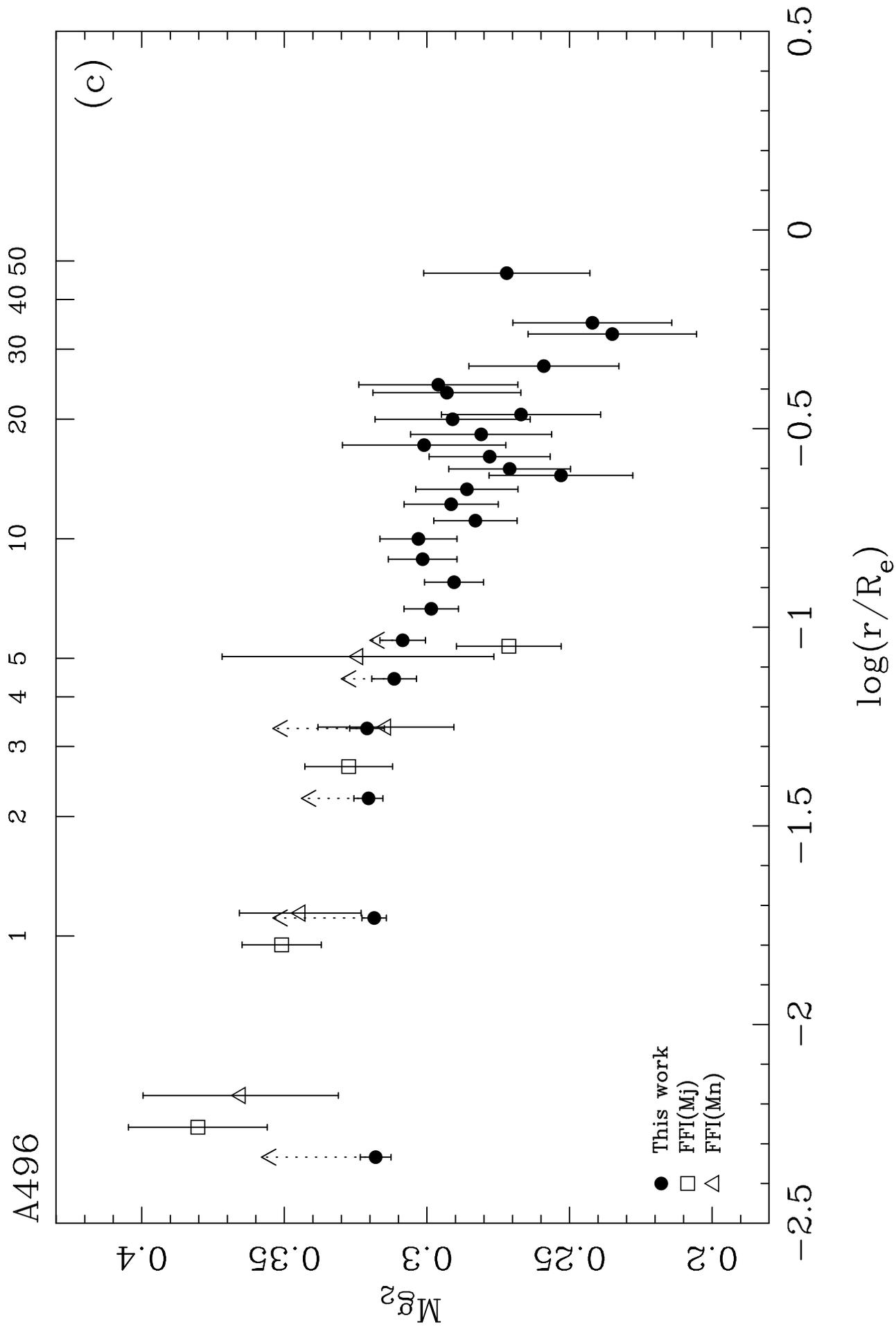
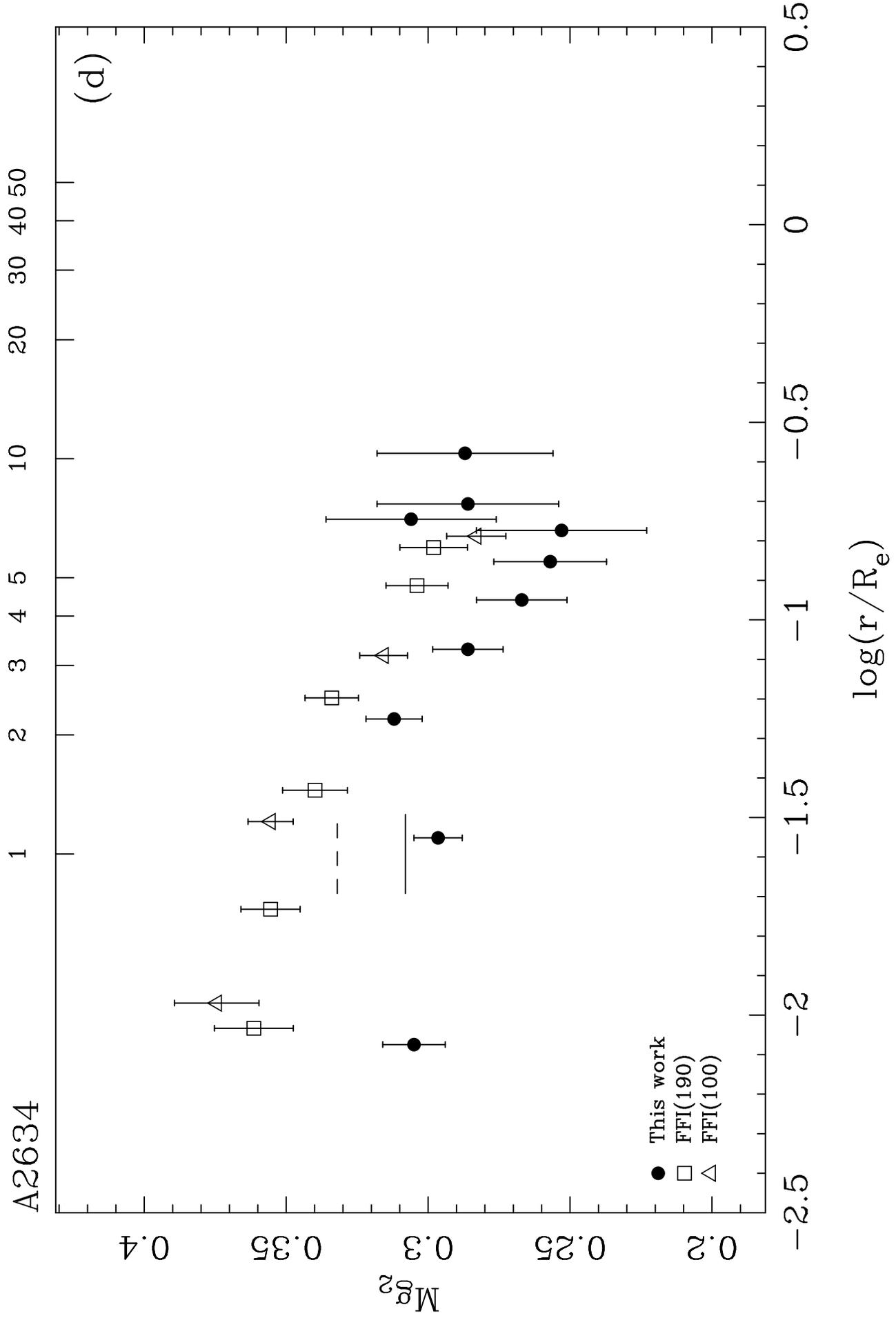
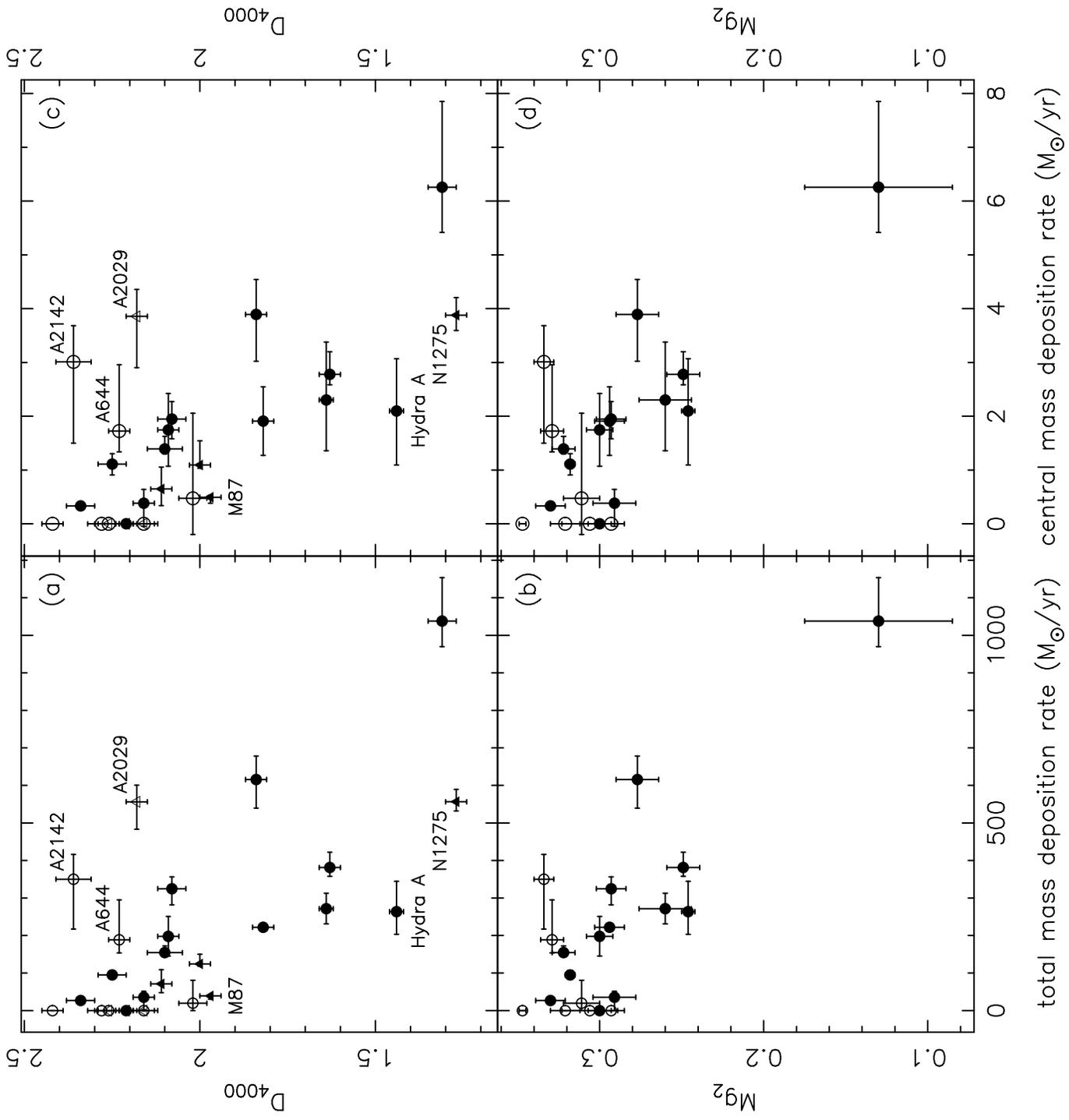
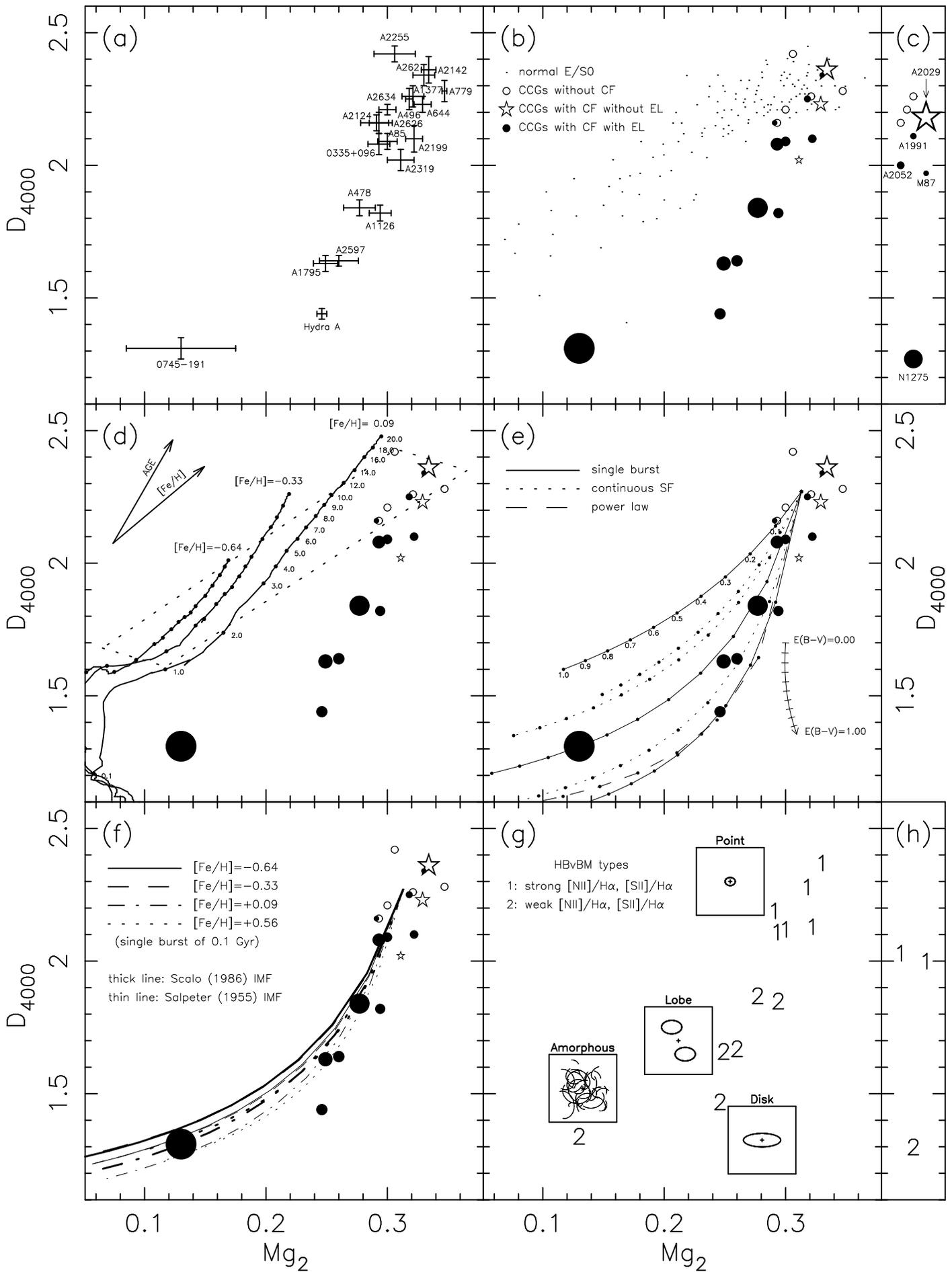
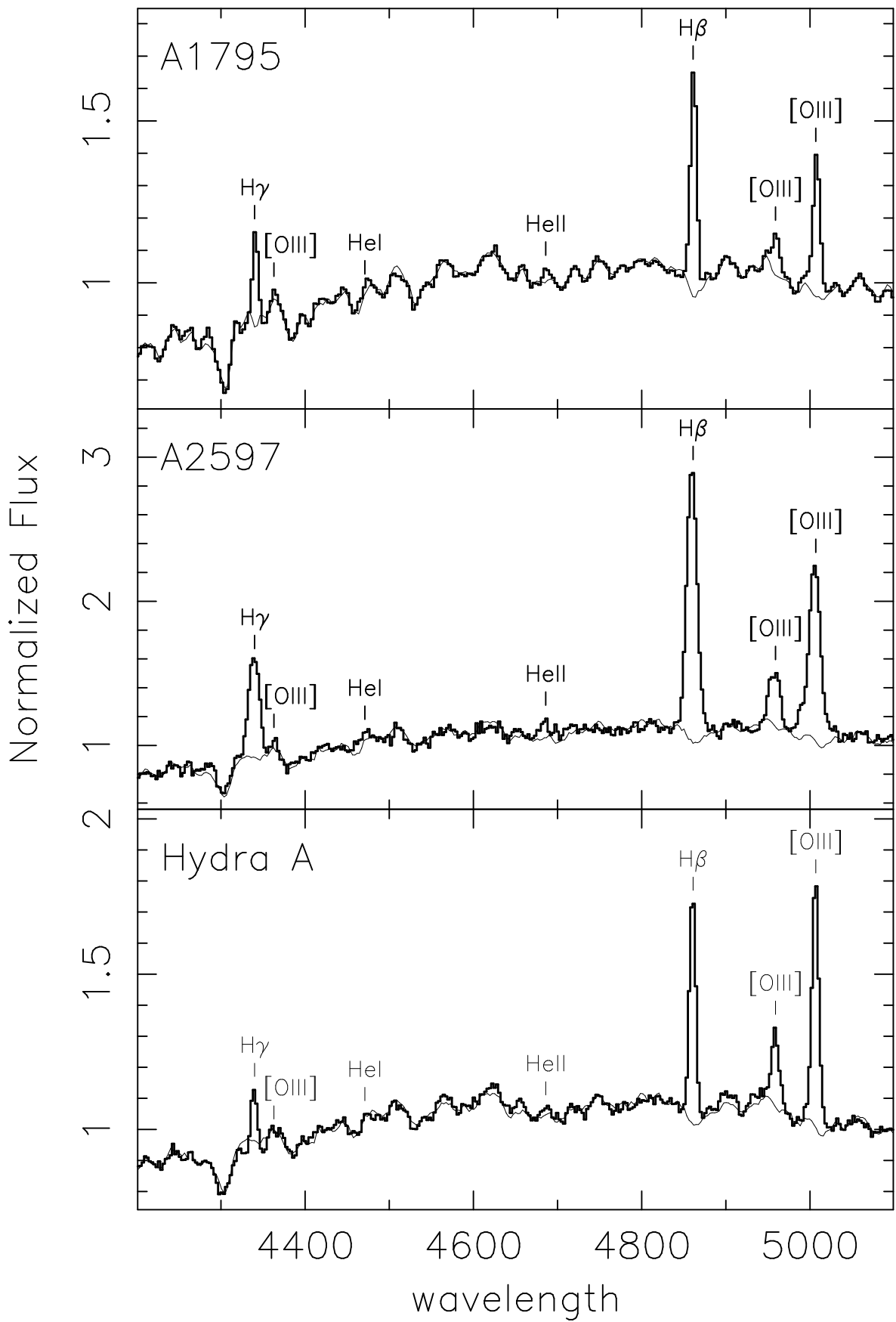
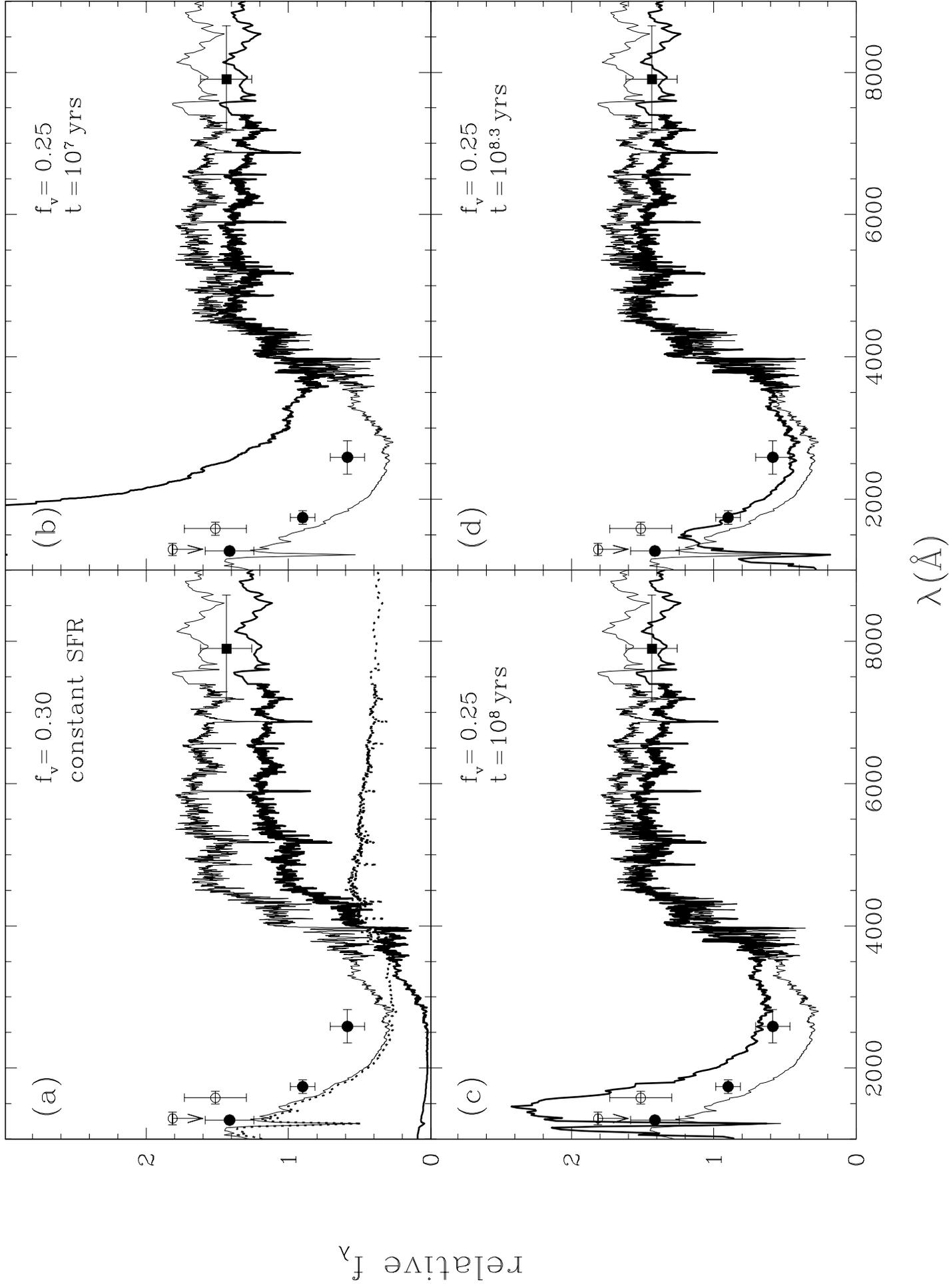
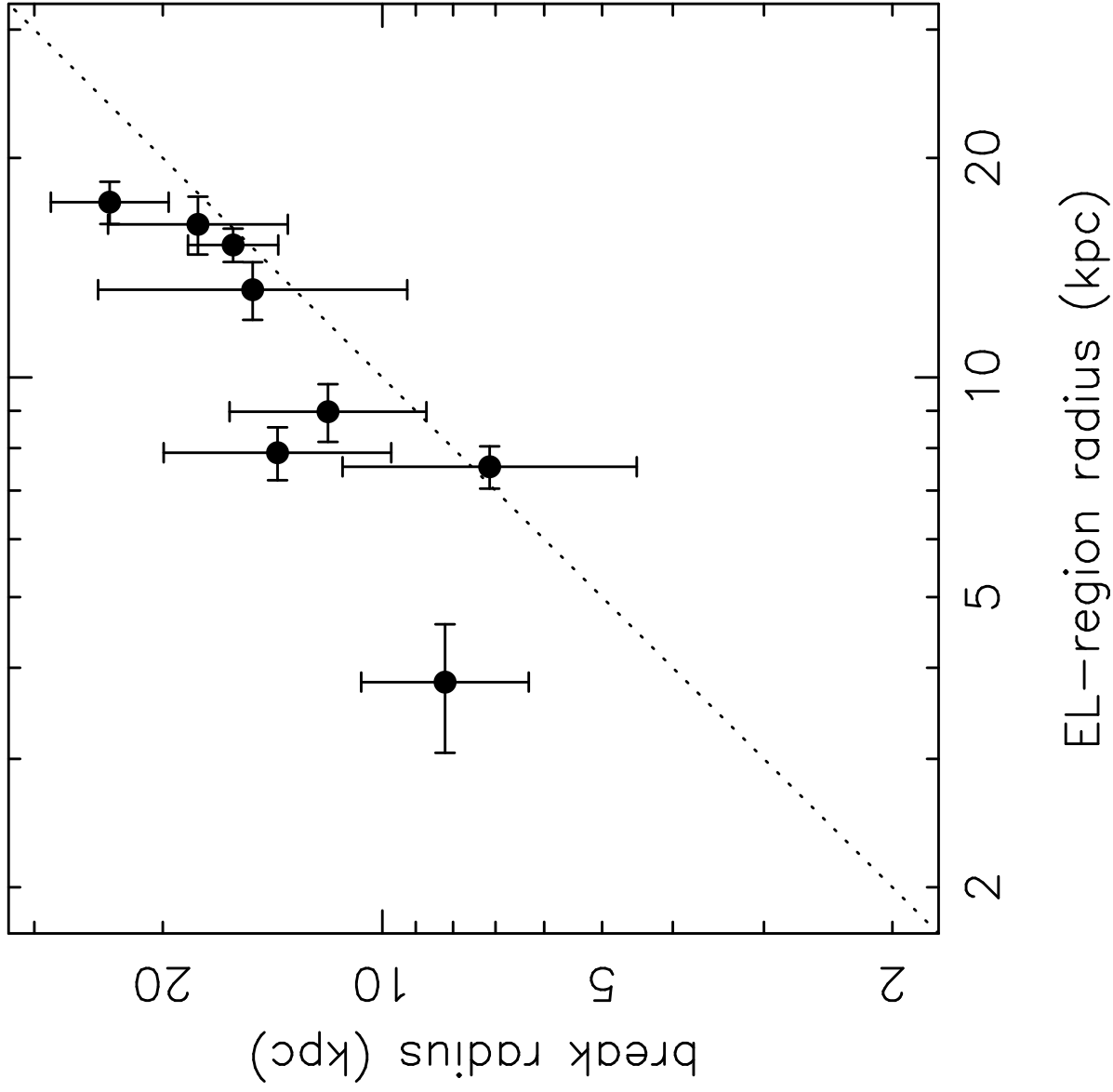
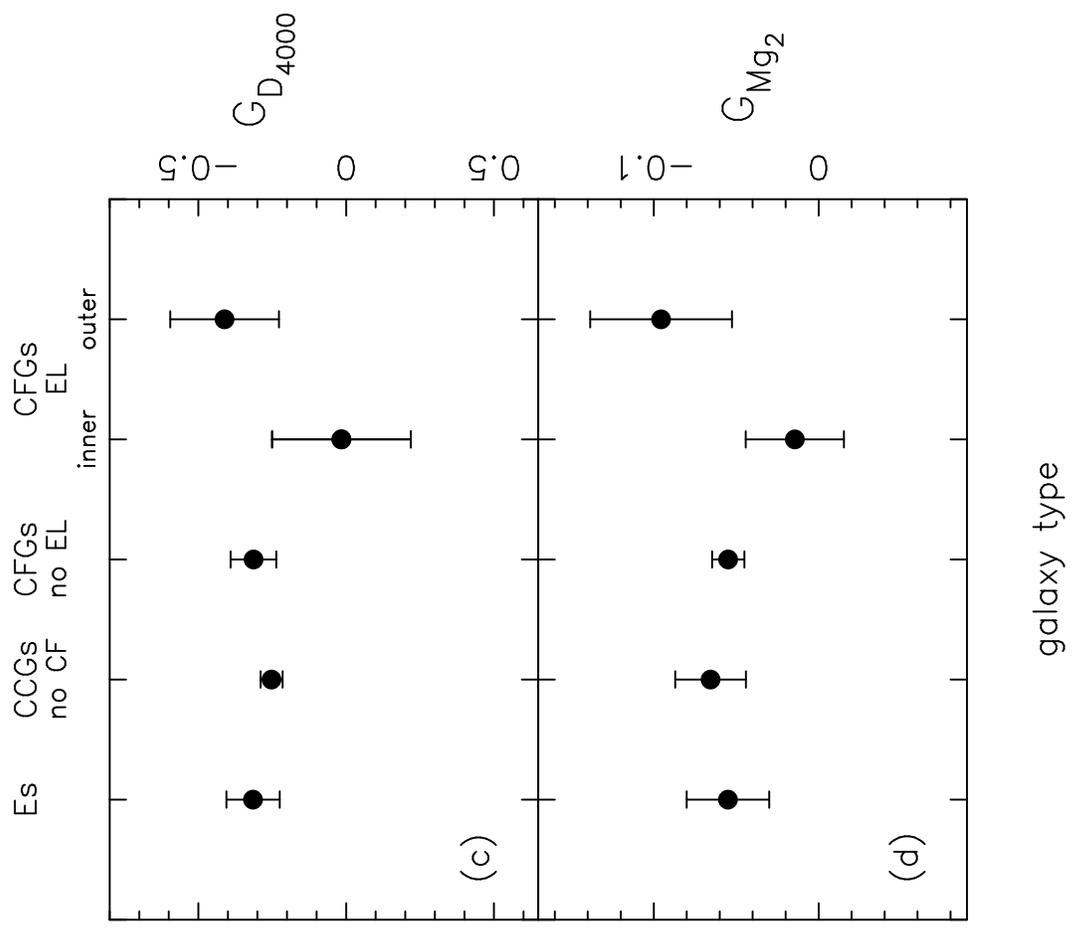
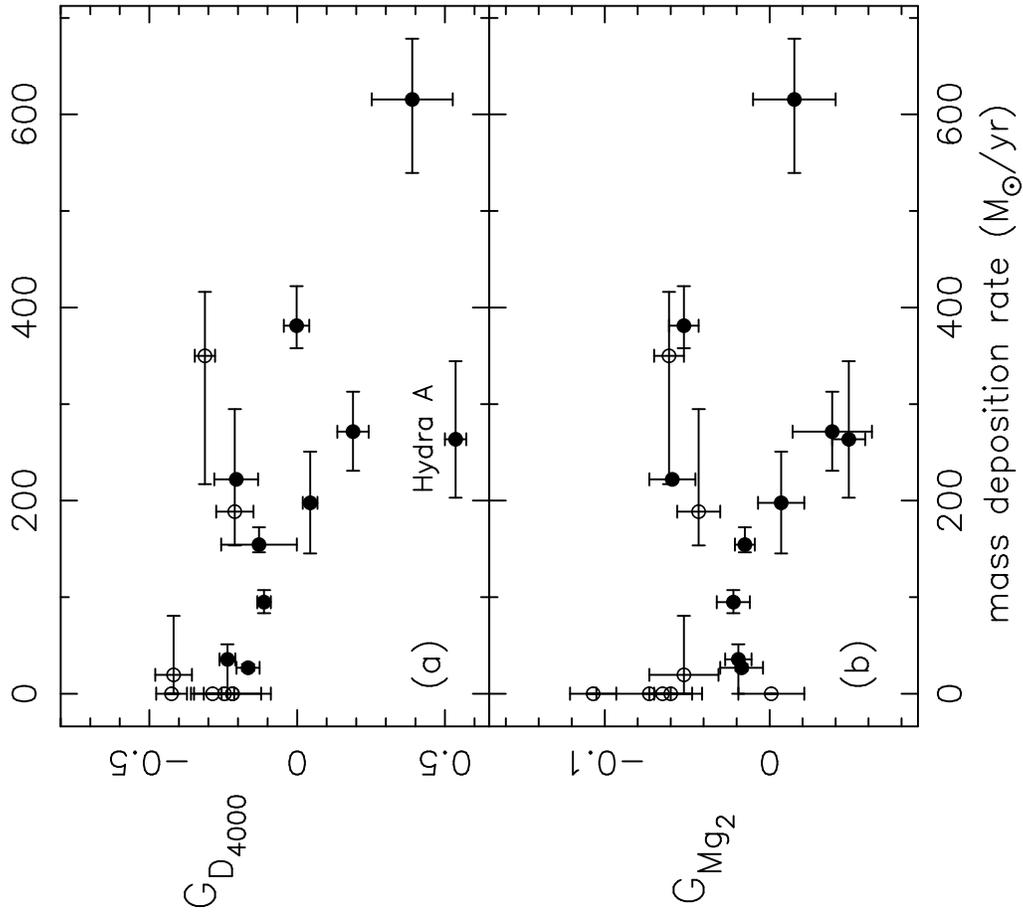
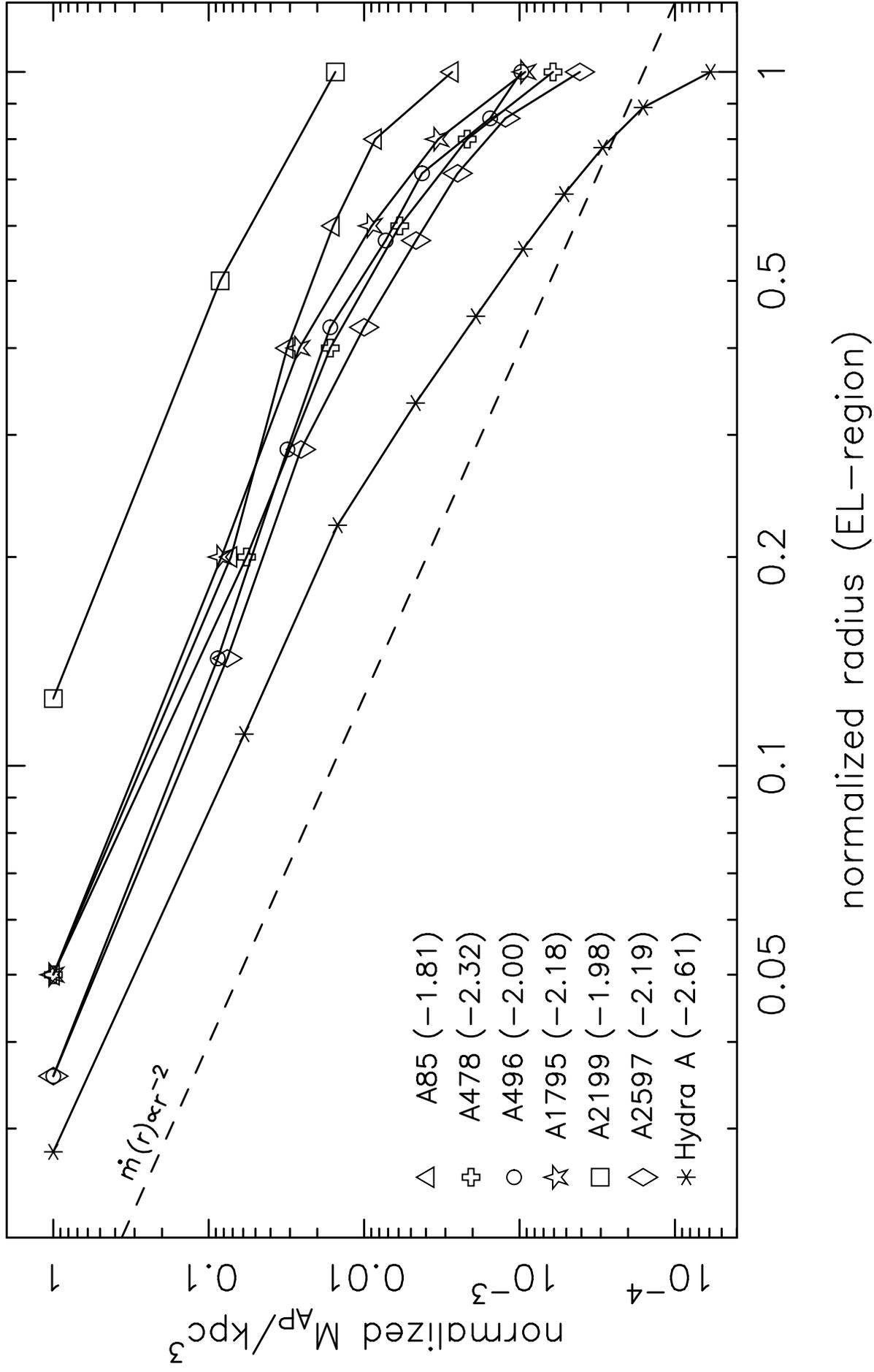
\end{document}